\documentclass[twocolumn]{aastex631}
\usepackage{longtable}

\usepackage[utf8]{inputenc}
\usepackage[flushleft]{threeparttable}
\usepackage{longtable}
\usepackage{gensymb}
\usepackage{amssymb}

\usepackage{multirow}
\usepackage{array}
\usepackage{amsmath}
\usepackage{mathtools}

\shorttitle{SNe II light curves}
\shortauthors{Fang et al.}

\begin{document}
%\begin{CJK*}{UTF8}{gbsn}

% Title.
%%%%%%%%%%%%%%%%%%%%%%%%%%%%%%%%%%%%%%%%%%%%%%%%%%%%%%%%%%%%%%%%%%%%%%%%%%%%%%%%%
\title{Diversity in hydrogen-rich envelope mass of type II supernovae (I): Plateau phase light curve modeling}
% Authors.
\author[0000-0002-1161-9592]{Qiliang Fang}
\affiliation{National Astronomical Observatory of Japan, National Institutes of Natural Sciences, 2-21-1 Osawa, Mitaka, Tokyo 181-8588, Japan}
\affiliation{Department of Astronomy, Kyoto University, Kitashirakawa-Oiwake-cho, Sakyo-ku, Kyoto 606-8502, Japan}
\author[0000-0003-2611-7269]{Keiichi Maeda}\affiliation{Department of Astronomy, Kyoto University, Kitashirakawa-Oiwake-cho, Sakyo-ku, Kyoto 606-8502, Japan}
\author[0009-0002-0706-1183]{Haonan Ye}\affiliation{Department of Astronomy, Kyoto University, Kitashirakawa-Oiwake-cho, Sakyo-ku, Kyoto 606-8502, Japan}
\author[0000-0003-1169-1954]{Takashi J. Moriya}
\affiliation{National Astronomical Observatory of Japan, National Institutes of Natural Sciences, 2-21-1 Osawa, Mitaka, Tokyo 181-8588, Japan}
\affiliation{Graduate Institute for Advanced Studies, SOKENDAI, 2-21-1 Osawa, Mitaka, Tokyo 181-8588, Japan}
\affiliation{School of Physics and Astronomy, Monash University, Clayton, Victoria 3800, Australia}
\author[0000-0002-9350-6793]{Tatsuya Matsumoto}
\affiliation{Department of Astronomy, Kyoto University, Kitashirakawa-Oiwake-cho, Sakyo-ku, Kyoto 606-8502, Japan}
\affiliation{Hakubi Center, Kyoto University, Yoshida-honmachi, Sakyo-ku, Kyoto 606-8501 Japan}

% Abstract
\begin{abstract}
We present a systematic study of Type II supernovae (SNe II) originating from progenitors with effective temperatures ($T_{\rm eff}$) and luminosities closely resembling red supergiants (RSGs) observed in pre-SN images and in the Galaxy. Using $\texttt{MESA}$, we compute a large grid of massive stars with $T_{\rm eff}$ ranging from 3200 K to 3800 K at their RSG phases, with hydrogen envelopes artificially stripped to varying extents (3 to 10\,$M_{\odot}$). The light curves of SNe IIP resulting from the explosions of these Galactic-RSG-like progenitors are modeled using $\texttt{STELLA}$. Our survey of the light curves reveals that partial stripping of the hydrogen envelope creates diversity in the magnitude and duration of SNe IIP light curves, without affecting the position of the RSG progenitor on the Hertzsprung-Russell diagram (HRD). For these Galactic-RSG-like progenitor models, we establish an indicator based on the light-curve properties to estimate the hydrogen envelope mass. Additionally, we discuss the effects of material mixing and $^{56}$Ni heating. Applying our model grid to a large sample of approximately 100 observed SNe IIP reveals a considerably broader range of hydrogen-rich envelope masses than predicted by standard stellar wind models. This finding suggests that, if SNe IIP are explosions of Galactic-like RSGs, to explain the diversity in the observed light curves, a significant fraction of them must have experienced substantial mass loss beyond the standard mass-loss prescription prior to their explosions. This finding highlights the uncertainties involved in massive star evolution and the pre-SN mass-loss mechanism.
\end{abstract}
 
% Official ApJ keywords in alphabetical order.
% See http://www.journals.uchicago.edu/ApJ/information.html

%%%%%%%%%%%%%%%%%%%%%%%%%%%%%%%%%%%%%%%%%%%%%%%%%%%%%%%%%%%%%%

\section{Introduction}
Core-collapse supernovae (CCSNe) are catastrophic explosions that are believed to occur in massive stars (typically with ZAMS mass $M_{\rm ZAMS}\, \gtrapprox 8\,M_{\rm \odot}$) once the fuel in their cores is exhausted. CCSNe exhibit a wide range of observable characteristics, and a primary goal of modern stellar physics is to establish a connection between this diversity and the massive progenitor stars that give rise to them.

Type II supernovae (SNe II), which are the most commonly observed CCSNe, show hydrogen features in their spectra, indicating the presence of a massive hydrogen-rich envelope in their progenitors (\citealt{filippenko97,galyam17,modjaz19}). SNe II are characterized by the plateau phase in their light curves. During this phase, the brightness remains almost constant for approximately 50 to 100 days due to the recombination of the hydrogen in the envelope. Following the expansion of the ejecta, the photosphere gradually descends inward and finally reaches to the bottom of the hydrogen-rich envelope, resulting in the sudden drop of the light curve brightness. The ejecta then enters so called the nebular phase. Pre-explosion photometry confirms the red supergiants (RSGs) as progenitors for a limited number of SNe II, where the ZAMS masses of these progenitors are suggested to be $\lessapprox$\,17\,$M_{\rm \odot}$ (\citealt{vandyk03, vandyk12a, vandyk12b, vandyk19, vandyk23a, vandyk23b, smartt04, maund05a, maund05b, maund13, maund14a, maund14b, li06, fraser10, fraser11, fraser12, fraser14, crockett11, eliasrosa11, kochanek12, kochanek17, tomasella13, oneill19, rui19, sollerman21, jencson23, kilpatrick23}). The lack of progenitors with $M_{\rm ZAMS}$\,$>$\,17\,$M_{\rm \odot}$, known as the RSG problem, presents a challenge (\citealt{smartt09, smartt15, walmswell12, eldridge13, meynet15, davies18, hiramatsu21, strotjohann23}). However, determining $M_{\rm ZAMS}$ based on pre-explosion photometry is often uncertain due to various factors such as the lack of multi-band photometry, the uncertainties in reddening estimates, and the limitations of stellar evolution models. Further, the progenitor before the explosion is usually too dim even for deep imaging observations, making pre-SN photometry only feasible for a limited number of SNe II (\citealt{smartt09}).

Radiation hydrodynamics and radiative transfer calculations are frequently utilized to constrain the nature of SNe II in the literature (See for example \citealt{snec15,martinez20}). This approach involves the evolution of the progenitor models with varying $M_{\rm ZAMS}$ up to the onset of core-collapse, followed by the deposition of energies into their cores to trigger the explosions. Sometimes non-evolutionary progenitor models are employed. The initial density and chemical composition distributions, along with their masses and radii, of these models are set as free parameters (\citealt{bersten11, martinez19}). Following the construction of the progenitor models, radioactive $^{56}$Ni is manually introduced into the ejecta, with variations in their amounts and distributions. The light curves of these models are then calculated and compared with those from observation, which allows us to extra the properties of the progenitor and the nature of the explosion (\citealt{morozova16,morozova17,morozova18,martinez22a,martinez22b,martinez22c,subrayan23,zha23}).

While this approach is frequently applied, it has several limitations. The plateau phase of the light curve is driven by the recombination of hydrogen in the envelope, making its characteristics primarily determined by the properties of the hydrogen-rich region (\citealt{popov93,kasen09,dessart19}). Inferring $M_{\rm ZAMS}$ from light curve modeling relies on the underlying assumption of the unique relation between the properties of the envelope and $M_{\rm ZAMS}$. However, there are several uncertainties involved in establishing this relation: (1) The formula that describes the RSG mass-loss rate, which is a function of the star's properties (e.g., mass, radius, luminosity, metallicity), is empirically derived from observation and involves many uncertainties in both observation and theory (\citealt{reimers75,lamers81,dejager88,nieuwenhuijzen90,kudritzki00,nugis00a,nugis00b,willson00,vink01,maeder01,schroder05,vanloon05,eldridge06,beasor20,vink23}).  
The accuracy of the RSG mass-loss rate when applied to the progenitors of SNe II therefore remains uncertain; (2) massive stars can be born in binary systems, where the amount of envelope stripping is primarily determined by the orbital parameters such as the mass ratio or the separation between the primary and secondary stars. In this case, the dependence of envelope mass on $M_{\rm ZAMS}$ becomes invisible (\citealt{heger03, eldridge08, yoon10, smith11, sana12, groh13, smith14, yoon15, yoon17, ouchi17, eldridge18, fang19, gilkis22, chen23, drout23, ercolino23, fragos23, hirai23, matsuoka23, sun23, dessart24}, among many others). 
 
To illustrate the uncertainties discussed above, \citet{dessart19} calculated the light curves of a grid of progenitor models with the same envelope mass ($M_{\rm Henv}\,\sim$\,8\,$M_{\rm \odot}$) and explosion energy (1.25\,$\times$\,10$^{51}$\,erg) but different $M_{\rm ZAMS}$ (12 to 25\,$M_{\rm \odot}$), and the light curves at the plateau phase were found to be similar for all models. \citet{goldberg19} also revealed that models with varying ejecta masses $M_{\rm eje}$ can produce similar light curves ($M_{\rm eje}$ includes the masses of the material below the hydrogen-rich envelope, which can be comparable to $M_{\rm Henv}$ if $M_{\rm ZAMS}$ is large). These works highlight the non-uniqueness of progenitor properties inferred from light-curve modeling.

In this work, we extend the analysis outlined in \citet{dessart19}. We evolve progenitor models with $M_{\rm ZAMS}$ ranging from 10 to 20\,$M_{\rm \odot}$, manually removing the envelope mass $M_{\rm Henv}$ to 3\,$\sim$\,12\,$M_{\rm \odot}$. Our progenitor models are constructed under a constraint that they should have $T_{\rm eff}$ between 3200 and 3800 K, in order to be consistent with RSGs observed in the Galaxy and estimates derived from pre-SN images of SN IIP RSG progenitors. The progenitor models are then exploded by manually injecting varied amounts of energies, and the corresponding light curves are calculated. Based on the survey of this light curve model grid, we conclude that, the light curve characteristics of the explosion of Galactic-like RSGs contain little information on $M_{\rm ZAMS}$, but are mainly affected by $M_{\rm Henv}$. Light curve modeling, in the absence of prior knowledge regarding the mass-loss history, does not provide informative constrains on the $M_{\rm eje}$ or $M_{\rm ZAMS}$. However, it does allow a precise estimation of the envelope mass within an uncertainty of 1\,$M_{\rm \odot}$. The inferred distribution of the envelope masses for a sample of SNe II reveals a considerably broader range compared to the predictions of single star models evolving with the standard stellar-wind prescription. This inconsistency highlights the uncertainty involved in the mass-loss history before the explosion.

This paper is organized as follow. In \S 2, we introduce the numerical approach, including the evolution of a grid of progenitors and their light curves. In \S 3, we present the sample survey for the obtained light curves, and establish the scaling relations between the characteristics of the light curves (plateau duration and magnitude) and the hydrogen-rich envelope mass, the radius and explosion energy for models without $^{56}$Ni, for the progenitor models adopted in the present work. Based on these relations, we establish a method to constrain the envelope mass within uncertainty of 1\,$M_{\rm \odot}$. The effects of the $^{56}$Ni heating and material mixing are also discussed. In \S 4, the results from \S 3 are applied to observational data. We derive the distribution of hydrogen-rich envelope masses $M_{\rm Henv}$ for a large sample of SNe II ($N\,\sim\,$100), and discuss its implications for massive star evolution and pre-SN mass-loss mechanism(s). \S 5 discusses the factors that may affect the results in \S 3. The conclusions are presented in \S 6.

\section{Numerical setup}

\subsection{Progenitor calculation}
The SN progenitor models are constructed using the one-dimensional stellar evolution code, Modules for Experiments in Stellar Astrophysics ($\texttt{MESA}$, \citealt{paxton11, paxton13, paxton15, paxton18, paxton19, mesa23}). We start with $\texttt{MESA}$ version r22.11.1 test suite $\texttt{example\_make\_pre\_ccsn}$ to create non-rotating, solar metallicity progenitor models. The ZAMS masses $M_{\rm ZAMS}$ are selected to be 10, 12, 15, 18 and 20\,$M_{\rm \odot}$, which encompass the typical mass range for SNe II progenitors (see for example \citealt{smartt15,valenti16,davies18,davies20}).
For the fiducial models, we employ the same mixing scheme as \citet{martinez20}, i.e., Ledoux criterion for convection, mixing length parameter $\alpha_{\rm MLT}\,$=\,2.0, exponential overshooting parameters $f_{\rm ov}$\,=\,0.004 and $f_{\rm ov, 0}$\,=\,0.001, semiconvection efficiency $\alpha_{\rm sc}$\,=\,0.01 (\citealt{farmer16}) and thermohaline mixing coefficient $\alpha_{\rm th}$\,=\,2 (\citealt{kippenhahn80}).  We additionally evolve another model grid with $\alpha_{\rm MLT}\,$=\,2.5, $M_{\rm ZAMS}$\,=\,10, 12, 15 and 18\,$M_{\rm \odot}$, with other parameters kept fixed. 
However, in this work, except for models evolved with the $\texttt{Dutch}$ scheme \citealt{dejager88,vink01,glebbeek09} and wind efficiency $\eta$\,=\,1.0, we consider the hydrogen-rich envelope mass $M_{\rm Henv}$ as a free variable to account for the uncertainties in the mass-loss mechanism such as stable/unstable binary mass-transfer or violent activity of massive stars (see, e.g., \citealt{smith14} for a review). Rather than self-consistently modeling these complicated processes, we evolve the progenitors models from pre-ZAMS to the depletion of helium in the core without wind mass-loss ($\eta$ = 0.0), and subsequently use the command $\texttt{relax\_mass\_to\_remove\_H\_env}$ to artificially remove the hydrogen-rich envelope, with maximum mass-loss rate held constant at 10$^{-2}~M_{\rm \odot}$\,yr$^{-1}$ ($\texttt{lg\_max\_abs\_mdot}$\,=\,-2). The stripped models are then further evolved to the depletion of carbon in the core fixing $\eta$ = 0.0.  Our own experiment shows that varying $\texttt{lg\_max\_abs\_mdot}$\, in a range between -1 and -4 will not affect the final radius. The residual $M_{\rm Henv}$ is controlled by the command $\texttt{extra\_mass\_retained\_by\_remove\_H\_env}$ and ranges from 3 to 14\,$M_{\rm \odot}$ in steps of 1\,$M_{\rm \odot}$ (the upper limit of the residual envelope mass depends on $M_{\rm ZAMS}$). We note here that, after the stripping procedure, the subsequent carbon burning phase will slightly increase the helium core mass by 0.02 to 0.1\,$M_{\rm \odot}$, and $M_{\rm Henv}$ is slightly decreased according, so the final $M_{\rm Henv}$ is not exactly the same as $\texttt{extra\_mass\_retained\_by\_remove\_H\_env}$. However, such difference is small. 

With these setups, the progenitor models are evolved from pre-main-sequence to the depletion of carbon in the core. The models are not evolved to the core collapse in this study for the reasons below: (1) For models with $M_{\rm ZAMS}$\,$\le$\,12\,$M_{\rm \odot}$, the calculation of the advanced fusions becomes computationally expensive and time-consuming. Some of the models develop off-center flames, leading to convergence problem during core oxygen burning phase. In fact, the products of the advanced fusions are mostly excised when the explosions are phenomenologically triggered (\S 2.2) and are not relevant to this study; (2) this study focuses on the plateau phase of SNe II light curves, which is primarily determined by the explosion energy and the properties of the hydrogen-rich envelope. After the carbon depletion, the outermost envelope is detached from the subsequent core evolution. A detailed discussion on this topic is deferred to \S 5.2. 

\begin{figure}[!htb]
\epsscale{1.15}
\plotone{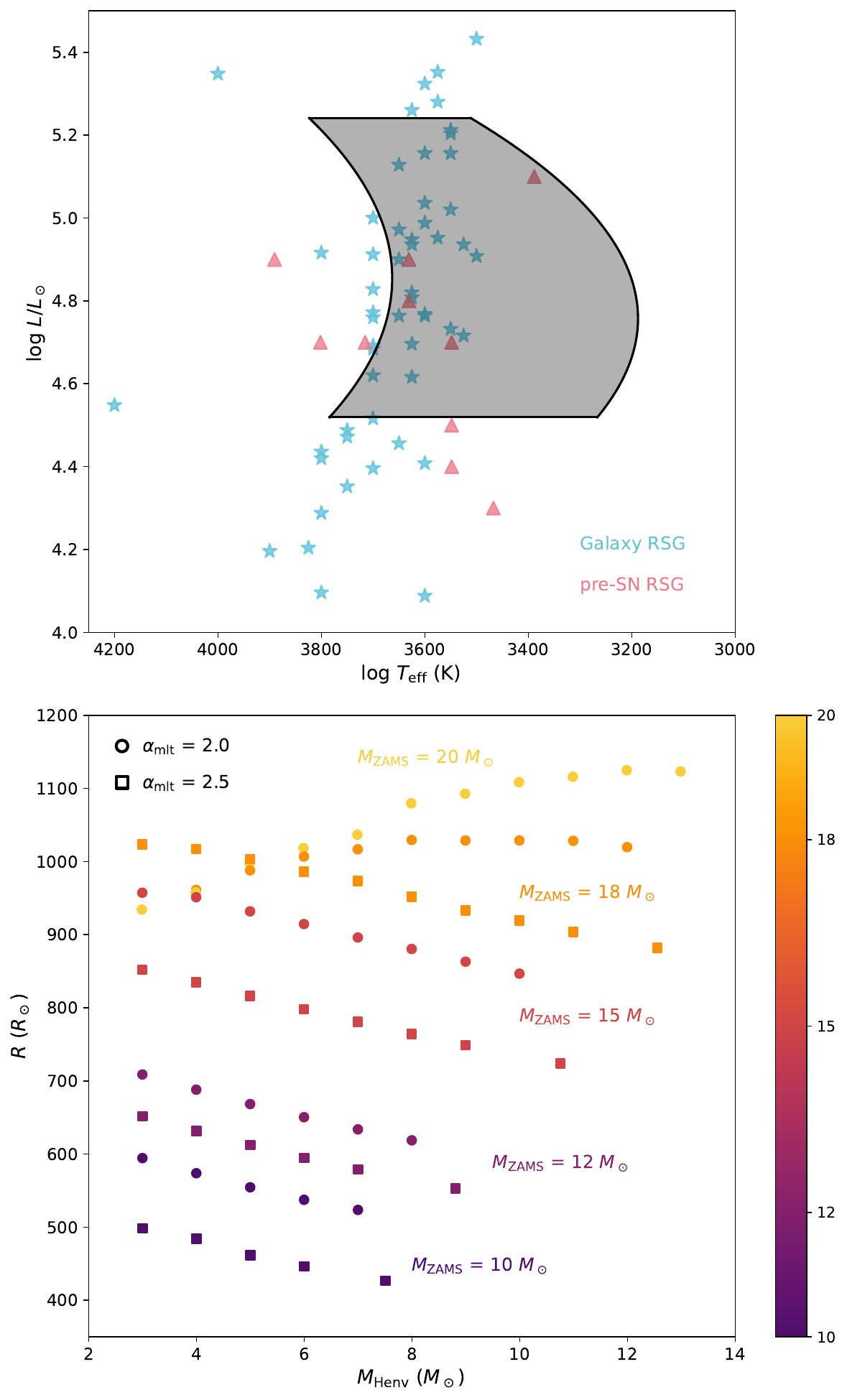}
\centering
\caption{Upper panel: The shaded region marks the range of progenitor models on HRD. The pink stars are RSGs from pre-SN images from \citet{smartt15}; The blue stars are RSGs in the Galaxy from \citet{levesque05}; Lower panel: The hydrogen-rich envelope mass $M_{\rm Henv}$ and the radius $R$ of the progenitor models in this work. Individual models are color coded by $M_{\rm ZAMS}$. Different markers represent models with different $\alpha_{\rm mlt}$.}
\label{fig:progenitor}
\end{figure}

The progenitor properties at the point of the carbon depletion are summarized in Table~\ref{tab:progenitor}. In Figure~\ref{fig:progenitor}, the upper panel shows the range of the progenitor models on HRD, along with RSGs in the Galaxy and those detected from pre-SN images. In the lower panel, the radii $R$ and the hydrogen-rich envelope masses $M_{\rm Henv}$ are compared. In general, our progenitor models have effective temperature $T_{\rm eff}$ ranging from 3200 to 3800 K, similar to the RSGs in the Galaxy, and have radii ranging from 500 to 1100 $R_{\rm \odot}$, depending on $M_{\rm ZAMS}$ and $\alpha_{\rm mlt}$. The removal of the hydrogen-rich envelope will not significantly affect the radius and the helium core mass, therefore any effect associated with the envelope stripping considered here is not detectable in the pre-SN images.

\begin{deluxetable*}{ccccccc}[t]
\centering
\label{tab:progenitor}
\tablehead{
\colhead{$M_{\rm ZAMS}$}&\colhead{$\alpha_{\rm mlt}$}&\colhead{log\,$\frac{L}{L_{\rm \odot}}$}&\colhead{$T_{\rm eff}$}&\colhead{$M_{\rm Henv}$}&\colhead{$R$}&\colhead{$M_{\rm He\,core}$}
}
\startdata
10&2.0&4.55&3245\,-\,3458&3.0\,-\,7.0&523\,-\,594&2.63\\
&2.5&4.52&3482\,-\,3766&3.0\,-\,7.5&426\,-\,498&2.49\\
\hline
12&2.0&4.68&3206\,-\,3433&3.0\,-\,8.0&618\,-\,708&3.05\\
&2.5&4.72&3422\,-\,3716&3.0\,-\,8.8&552\,-\,651&3.19\\
\hline
15&2.0&4.94&3199\,-\,3394&3.0\,-\,10.0&846\,-\,957&4.24\\
&2.5&4.94&3389\,-\,3680&3.0\,-\,10.8&723\,-\,852&4.24\\
\hline
18&2.0&5.11&3414\,-\,3530&4.0\,-\,12.0&961\,-\,1029&5.59\\
&2.5&5.10&3395\,-\,3668&3.0\,-\,12.6&882\,-\,1023&5.44\\
\hline
20&2.0&5.21&3441\,-\,3857&3.0\,-\,13.0&934\,-\,1124&6.47\\
\enddata
\caption{Summary of the progenitor properties. Columns: ZAMS mass, $\alpha_{\rm mlt}$, luminosity of the RSG, $T_{\rm eff}$ of the RSG, hydrogen-rich envelope mass, stellar radius and helium core mass. The masses and radii are in solar unit. $T_{\rm eff}$ is in the unit of K.}
\end{deluxetable*}

\subsection{Hand-off to \texttt{STELLA}}
For handing-off the hydrostatic progenitor models to \texttt{STELLA} for light curve calculations, we closely follow the test suite $\texttt{ccsn\_IIp}$ to trigger the explosions. This simulation includes two procedures; the energy injection and the shock propagation. To perform the mass-cut that mimics the compact remnant formation, we specially select progenitor models with $M_{\rm ZAMS}$\,=\,12, 15, 18, 20\,$M_{\rm \odot}$, using \texttt{Dutch} wind scheme with $\eta$\,=\,1.0, and evolve these models to the onset of core-collapse, i.e., the point when iron core infall speed exceeds 100 km s$^{-1}$. The inner mass coordinate where entropy/baryon\,=\,4\,$k_{\rm B}$ are 1.4, 1.6, 1.9 and 2.0\,$M_{\rm \odot}$ respectively, which are subsequently selected as the mass-cuts for other models with the same $M_{\rm ZAMS}$ but different $M_{\rm Henv}$ (\citealt{ertl16}), as the loosely-attached envelope hardly affect the evolution of the inner core. For models with $M_{\rm ZAMS}$\,=\,10\,$M_{\rm \odot}$, the mass-cut is 1.2\,$M_{\rm \odot}$, aligned with the mass coordinate where entropy/baryon\,=\,4\,$k_{\rm B}$ when the model with $M_{\rm ZAMS}$ = 10\,$M_{\rm \odot}$ and $\eta$\,=\,1.0 is evolved to the core carbon depletion. As long as we are only interested in the plateau phase of the SNe II, the small variation in the mass-cut is indeed not important.

After the remnant is removed, the explosion energy is manually deposited in the inner 0.2\,$M_{\rm \odot}$ to induce the strong shock and trigger the explosion. A number of explosion models are calculated for each progenitor model,with the asymptotic energy (i.e., the energy stored in the expanding ejecta) ranging from 0.5 to 2.5\,$\times\,10^{51}$ erg (in 0.5\,$\times\,10^{51}$ erg steps). Hereafter we refer to 1\,$\times\,10^{51}$ erg as 1\,foe. For models with $M_{\rm ZAMS}$\,=\,10, 12 and 15\,$M_{\rm \odot}$, we additionally calculate explosions with the asymptotic energy of 0.3 foe. For models with $M_{\rm ZAMS}$\,=\,10 and 12\,$M_{\rm \odot}$, we further calculate low-energy events with the asymptotic energy = 0.1 foe. In the following text, we use the term `explosion energy' to refer to the asymptotic energy for convenience.

Strong shock is generated following the launch of the explosion, which then propagates through the envelope. During the shock propagation, the infalling material is removed by turning on the command $\texttt{fallback\_check\_total\_energy}$. After the shock front reaches to 0.05\,$M_{\rm \odot}$ below the stellar surface, the calculation is halted. At this point, we manually excise materials with fallback velocity larger than 500 km s$^{-1}$ and uniformly distribute radioactive $^{56}$Ni below the inner boundary of the hydrogen-rich envelope. The mass of $^{56}$Ni ($M_{\rm Ni}$) varies from 0.00 to 0.10\,$M_{\rm \odot}$ with 0.01\,$M_{\rm \odot}$ increments. Additional models with $M_{\rm Ni}$\,=\,0.001, 0.005, 0.008, 0.12 and 0.15\,$M_{\rm \odot}$ are also calculated. Both the amount and the distribution of $^{56}$Ni play roles in shaping the light curve characteristics (\citealt{kasen09,bersten11,moriya16}). Observationally, there is evidence that a fraction of $^{56}$Ni is mixed into the hydrogen-rich envelope. It has long been realized that substantial material mixing during the explosion is required to produce the observed smooth SNe II light curves. To mimic this effect, we apply the artificial `boxcar' averaging by setting the boxcar size to be 10\% of the helium core mass, which then runs through the ejecta for 4 times to average the mass fractions of the different elements (\citealt{kasen09,dessart12,dessart13,snec15}).  

With the above setups, the models are hand-off to $\texttt{STELLA}$, a one-dimensional multi-frequency radiation hydrodynamics code (\citealt{blinnikov98, blinnikov00, blinnikov06}), for the calculation of the light curves. We set 800 spatial zones and 40 frequency bins. No circumstellar material (CSM) is introduced. Models that take a long time to converge are simply discarded, as we are only interested in the bulk statistics of the model grid.

\section{Result}
In this section, we start with the sample survey of models without $^{56}$Ni, which serve as the reference models for the following discussions. We investigate how the basic parameters, i.e., the hydrogen-rich envelope mass $M_{\rm Henv}$, radius $R$, and the explosion energy $E$, affect the light curve characteristics. Especially, we focus on the duration and the magnitude of the plateau, and derive the scaling relations connecting these observables with the physical properties of the explosion. Based on these scaling relations, we establish a method to accurately constrain $M_{\rm Henv}$. The effects of the $^{56}$Ni heating on the light curve characteristics are also discussed. 

\subsection{Sample survey}
We first focus on models without $^{56}$Ni. For the model grid considered in this work, the duration of the plateau ranges from 40 to 120 days, with the peak magnitudes varying from 14.5 to 18.2 mag. The light curve characteristics are primarily determined by $M_{\rm Henv}$, $R$ and $E$. In general, large explosion energy and small hydrogen-rich envelope mass leads to bright and short plateau. While large radius increases the plateau magnitude, the duration is hardly affected.

\begin{figure}[!htb]
\epsscale{1.15}
\plotone{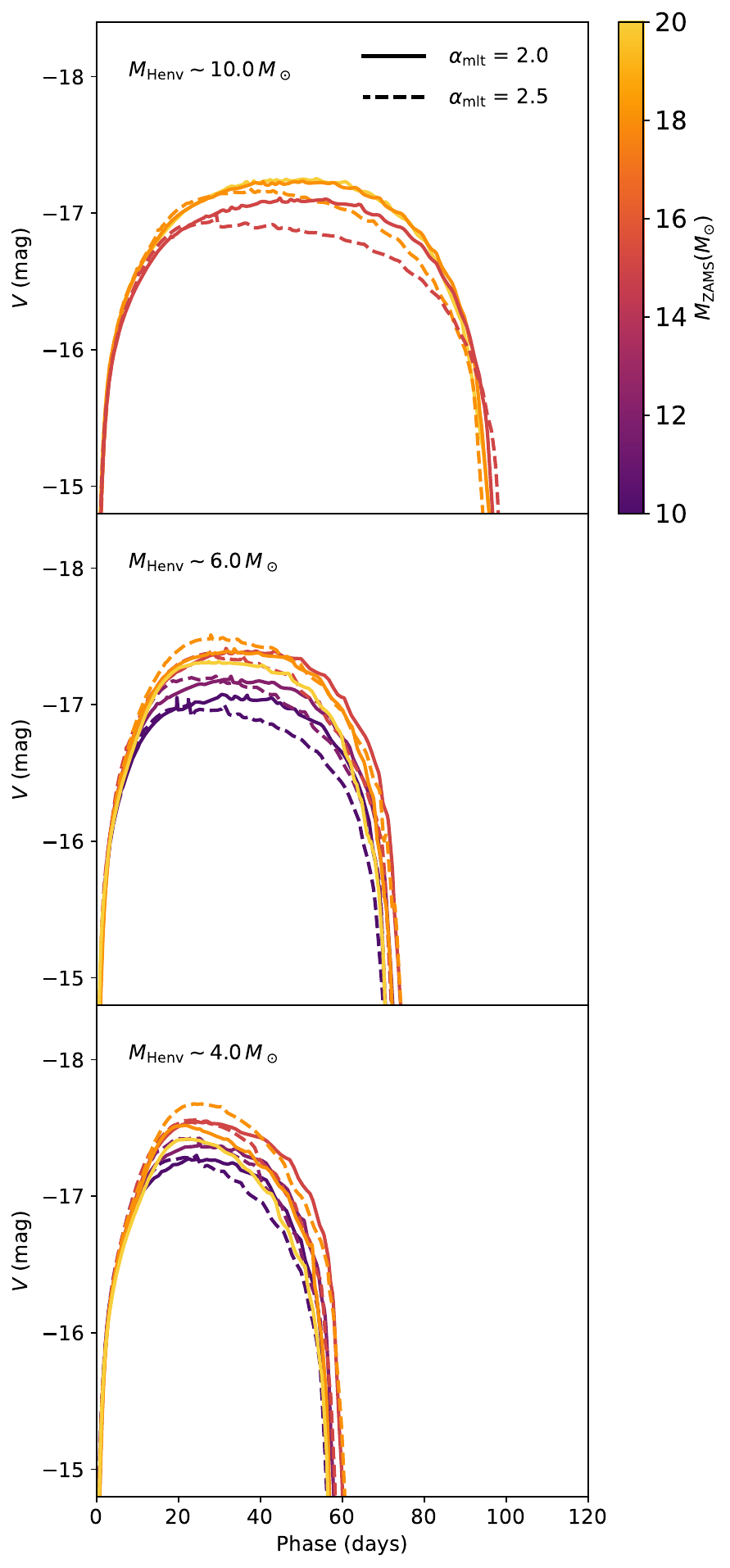}
\centering
\caption{The comparison between the $V$-band light curves of progenitors with different $M_{\rm ZAMS}$ but the similar $M_{\rm Henv}$ and the same $E$ (1 foe). From top to bottom panel: $M_{\rm Henv}$ = 10.0, 6.0, 4.0\,$M_{\rm \odot}$. The light curves of the models with $\alpha_{\rm mlt}\,$=\,2.0 and 2.5 are labled by solid and dashed lines respectively.}
\label{fig:lc_example}
\end{figure}

To be more specific, we compare the progenitor models with different $M_{\rm ZAMS}$ but similar $M_{\rm Henv}$. Figure \ref{fig:lc_example} shows the light curves of these models, all with a fixed explosion energy $E$\,=\,1\,foe. When $M_{\rm Henv}$ is fixed, the duration of the plateau is quite similar, while models with larger surface radii tend to exhibit brighter plateaus as expected (\citealt{popov93,kasen09,dessart13,dessart19}). Although models with larger $M_{\rm ZAMS}$ tend to be brighter when $M_{\rm Henv}$ and $E$ are kept fixed, such difference is too small to allow for unambiguous inference on the $M_{\rm ZAMS}$ of the progenitors, considering the uncertainties of distances and extinctions. The above discussion implies the light curves of SNe II at plateau phase provide limited information regarding $M_{\rm ZAMS}$ of their progenitors, and such degeneracy is valid for the typical range of $M_{\rm Henv}$ of SNe II (3 to 14\,$M_{\rm \odot}$; see for example \citealt{hiramatsu21}). 

In the literature, modeling the plateau phase light curve is a commonly adopted method to determine the ZAMS mass of the progenitor of SNe IIP (see for example \citealt{morozova18, martinez20}). while the SNe II light-curve characteristics, i.e., the duration and the magnitude of the plateau, are primarily determined by the properties of the hydrogen-rich envelope, rather than directly associated with the inner helium-core. Measuring $M_{\rm ZAMS}$ of the progenitor by light curve modeling thus relies heavily on the correlation between $M_{\rm Henv}$ and $M_{\rm ZAMS}$ predicted by the $standard$ wind mass-loss. However, the RSG mass-loss rates are not well constrained and the mass-loss mechanism (single star versus binary evolution) is not clear. There is thus no unique association developed between $M_{\rm Henv}$ and $M_{\rm ZAMS}$. A detailed discussion on this topic is deferred to \S 4.

\subsection{Scaling relations}
The scaling relations between the light-curve characteristics and the properties of the progenitor are useful to constrain the nature of SNe II. In the literature, ejecta mass ($M_{\rm eje}$), progenitor radius ($R$) and the explosion energy ($E$) are frequently employed as independent variables. Although other quantities, for example, opacity of the envelope ($\kappa$), ionization temperature of hydrogen ($T_{\rm I}$) or the helium fraction in the envelope ($X_{\rm He}$) are sometimes involved in deriving the scaling relations (\citealt{popov93, kasen09}), they are of secondary importance, and most of the analysis focus on $M_{\rm eje}$, $R$ and $E$ (\citealt{kepler16,goldberg19}).   

Based on the analytical model where the effects of radiative diffusion and hydrogen recombination are included, \citet{popov93} derived the scaling relations for the nickle-free models; 
\begin{equation} \label{Eq:popov93}
\begin{split}
V_{\rm p, 0} & \sim -1.67\,{\rm log}\,R+1.25\,{\rm log}\,M_{\rm eje}-2.08\,{\rm log}\,E \\
{\rm log}\,t_{\rm p,0} & \sim0.17\,{\rm log}\,R+0.57\,{\rm log}\,M_{\rm eje}-0.17\,{\rm log}\,E.
\end{split}
\end{equation}
Here, $V_{\rm p, 0}$ and $t_{\rm p,0}$ are the magnitude and duration of the plateau in $V$-band without radioactive heating. Here we only show the scaling because the constant terms vary between different works. \citet{kasen09} also reached to the similar results.

It is controversial as to which of the ejecta mass or the hydrogen-rich envelope mass should be used as an independent variable when applying the scaling relations of \cite{popov93}. For example, \citet{kepler16} employed $M_{\rm Henv}$, while \citet{goldberg19} suggested to use $M_{\rm eje}$ after finding some hydrogen elements are mixed deeply into the interior of the star due to Rayleigh Taylor instability (RTI). In a recent work, \citet{hsu24} find using $M_{\rm Henv}$ as a parameter can better characterize the scaling relations. In this work, we employ the artificial `boxcar' average to mimic large scale mixing, which is frequently adopted in SN II light curve modeling \citep{kasen09,snec15}. Based on this scheme, the hydrogen-rich envelope is only weakly mixed into the inner region and models with the same $M_{\rm Henv}$ have very similar light curves despite their large difference in $M_{\rm eje}$, as demonstrated in \S 3.1. We therefore adopt $M_{\rm Henv}$ rather than $M_{\rm eje}$ as the independent variable throughout this work.

We start with measuring the plateau magnitudes and duration of the $V$-band light curves in our model grid. There are several methods available to define the plateau duration, based on either theoretical models or observables. For example, \citet{goldberg19} defined $t_{\rm p}$ as the phase when the opacity of the inner boundary of the ejecta drops to $\tau$=10 ($t_{\rm \tau=10}$). Motivated by observation, \citet{valenti16} proposed to fit the light curves by the function 
\begin{equation} \label{Eq:valenti}
V(t) = \frac{A_0}{1 + e^{(t - t_{\rm p})/W_0}} + P_0 \times t + M_0. \\
\end{equation}
Here $t_{\rm p}$ defines the plateau duration, and other parameters together control the light-curve shape. The readers may refer to \citet{valenti16} for more details. However, this function requires the presence of a radioactive tail, and cannot produce reasonable fit to our reference models without $^{56}$Ni. We therefore employ a simple method to measure $t_{\rm p}$, which is determined by the phase when the $V$-band magnitude drops by 1 mag from the peak. We compare $t_{\rm p}$ measured in this way with $t_{\rm \tau=10}$, and find good agreement. In the following, the plateau duration $t_{\rm p}$ is defined in this way, and $t_{\rm p,0}$ represents the measurements for the $^{56}$Ni-free models.

In the literature, the magnitude (or luminosity) at 50 days after the shock breakout (SBO), $V_{\rm 50}$, is used to represent the plateau magnitude. However, for events with a very short plateau, $V_{\rm 50}$ is not well defined. In some extreme cases, the duration of the plateau is even shorter than 50 days. In this work, we measure the plateau magnitude $V_{\rm p}$ at $t$\,=\,0.5\,$\times\,t_{\rm p}$, i.e., the midpoint of the plateau. At this point, the magnitude is hardly affected by the presence of the circumstellar material (CSM) around the progenitor (\citealt{morozova17}; except for the case when the CSM is massive and extended) or by the $^{56}$Ni heating. Similarly to the definition of $t_{\rm p,0}$, $V_{\rm p,0}$ represents the measurements for the $^{56}$Ni-free models.

\begin{figure}[!htb]
\epsscale{1.15}
\plotone{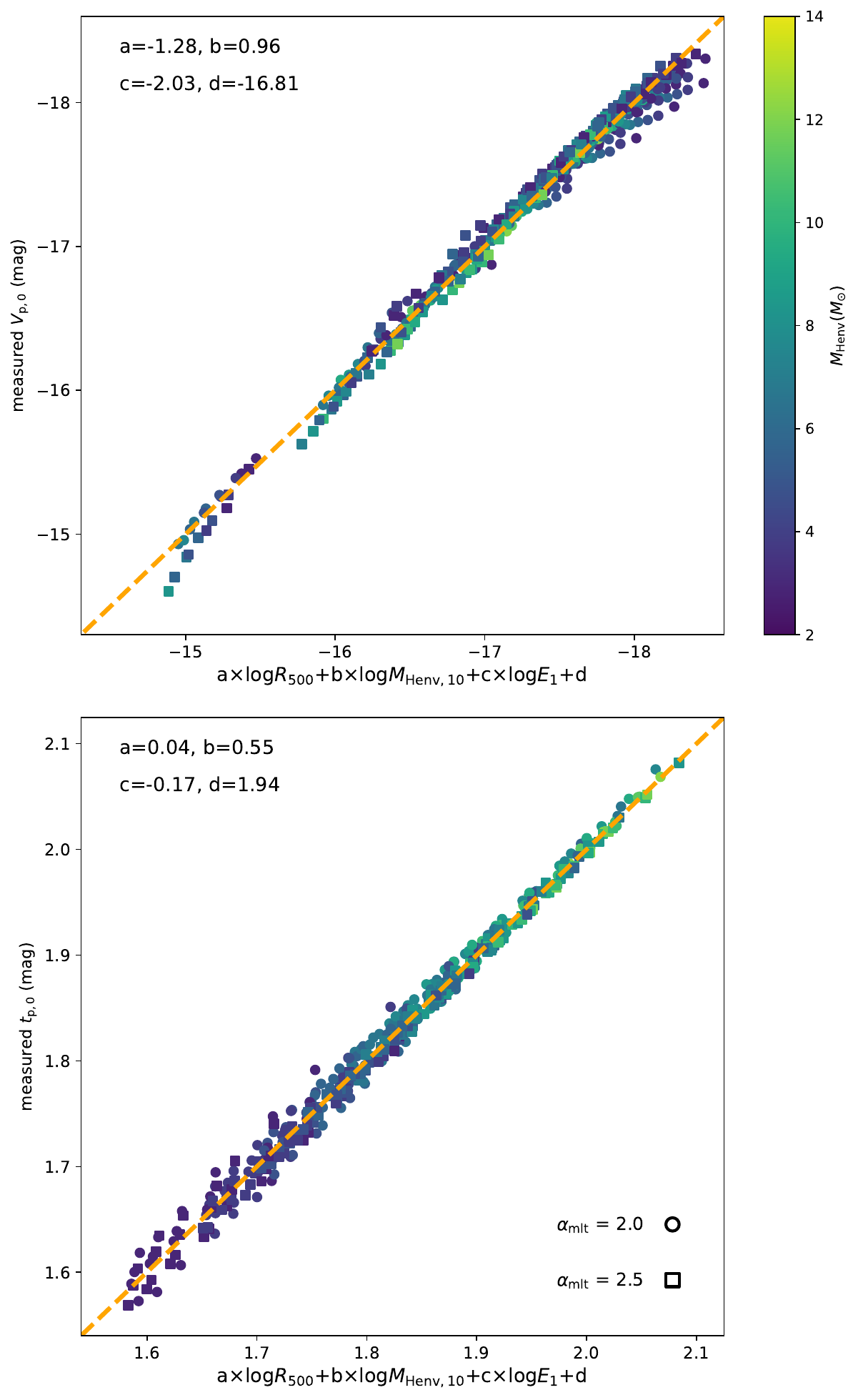}
\centering
\caption{The accuracy of Equation \ref{Eq:scaling_this_work} for plateau magnitude ($V_{\rm p,0}$, upper panel) and duration ($t_{\rm p,0}$, lower panel). Here, $R_{\rm 500}$, $M_{\rm Henv, 10}$ and $E_{1}$ are $R$, $M_{\rm Henv}$ and $E$ in the units of 500\,$R_{\rm \odot}$, 10\,$M_{\rm \odot}$ and 1 foe. The dashed line is one-to-one correspondence. The models are color coded by their $M_{\rm Henv}$, with gradual increase of $M_{\rm Henv}$ from the blue end to the green end.}
\label{fig:scaling}
\end{figure}

By fitting $V_{\rm p,0}$ and $t_{\rm p,0}$ with $M_{\rm Henv}$, $R$ and $E$ being independent variables, we establish the scaling relations for the models in this work as follows;
\begin{equation} \label{Eq:scaling_this_work}
\begin{split}
V_{\rm p, 0} & \sim -1.28\,{\rm log}\,R\,+\,0.96\,{\rm log}\,M_{\rm Henv}\,-\,2.03\,{\rm log}\,E \\
{\rm log}\,t_{\rm p,0} & \sim0.04\,{\rm log}\,R\,+\,0.55\,{\rm log}\,M_{\rm Henv}\,-0.17\,{\rm log}\,E.
\end{split}
\end{equation}
Figure \ref{fig:scaling} illustrates the accuracy of Equation \ref{Eq:scaling_this_work}. Notably, the dependence of $t_{\rm p,0}$ on $R$, as determined in this work (see also \citealt{lisakov17} based on \texttt{CMFGEN} modeling), is much weaker than that predicted by \citet{popov93} and \citet{kasen09}. 

As emphasized by \citet{kasen09}, the relation between the radius $R$ and the ejecta mass $M_{\rm eje}$ (or the hydrogen-rich envelope mass $M_{\rm Henv}$) predicted by the stellar evolution model naturally contributes to Equation \ref{Eq:scaling_this_work}. In this work, by artificially removing the hydrogen-rich envelope, we derive a grid of models with comparable $R$ but different $M_{\rm Henv}$ (Table~\ref{tab:progenitor} and Figure~\ref{fig:progenitor}). The relation between $R$ and $M_{\rm Henv}$ is eliminated, and the effects of these two quantities on the light curve characteristics are therefore constrained independently.

\subsection{Indicator of hydrogen-rich envelope mass}
From the previous sections, we show that the properties of SNe II light curves are primarily determined by the hydrogen-rich envelope and the explosion energy. Little information on the progenitor $M_{\rm ZAMS}$ can be extracted without a well-constrained mass-loss scheme. However, $M_{\rm Henv}$ itself is an important quantity that can be used to test the mass-loss scheme in massive star evolution (\S 4). In this section, our aim is to establish a measurement of $M_{\rm Henv}$ that can be applied to observation.

\begin{figure}[!htb]
\epsscale{1.15}
\plotone{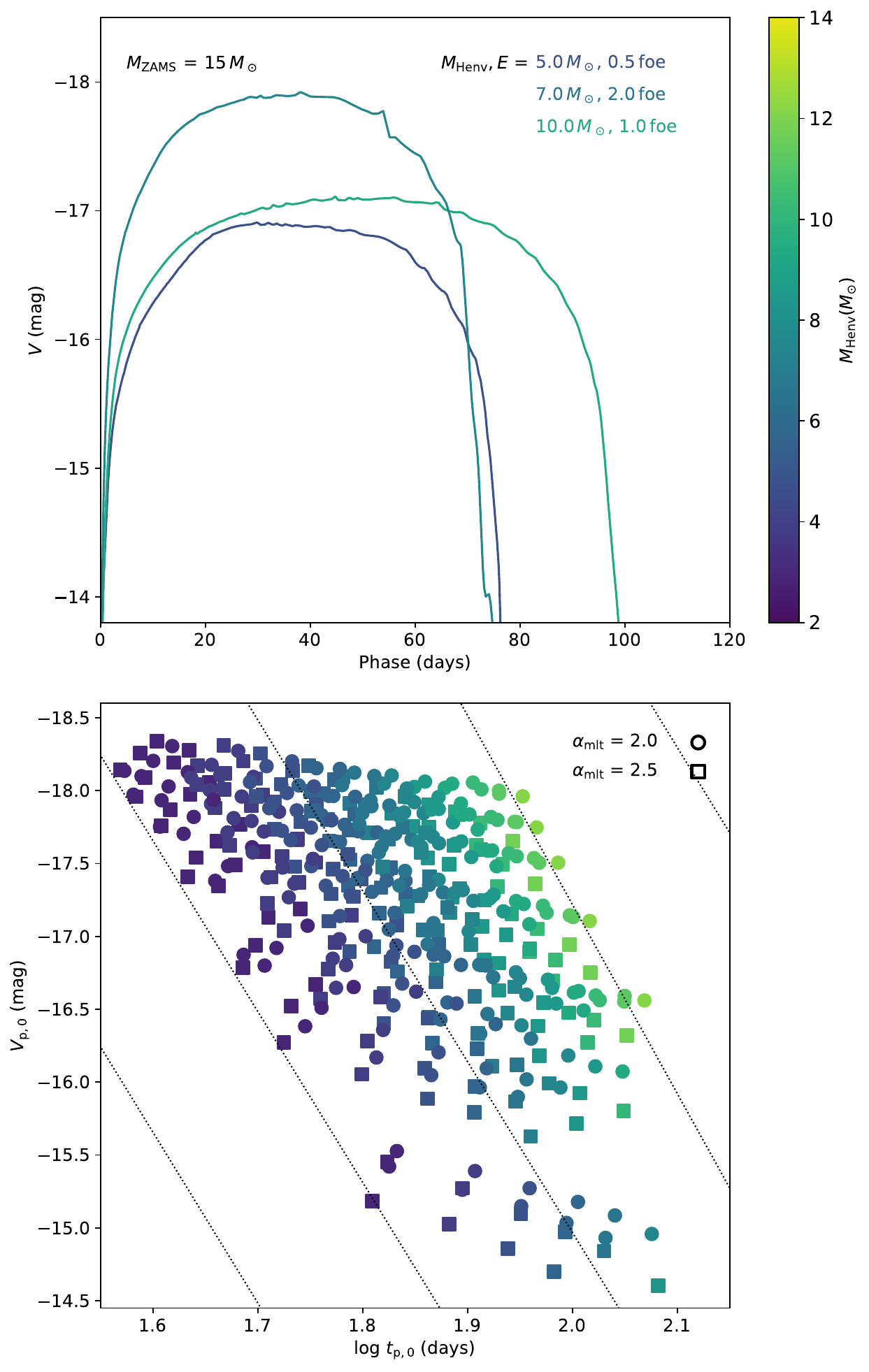}
\centering
\caption{Upper panel: The models with different $M_{\rm Henv}$, but with light curves adjusted to have the same $t_{\rm p,0}$ or $V_{\rm p,0}$. Lower panel: The relation between $t_{\rm p,0}$ and $V_{\rm p,0}$. The models are color coded by their $M_{\rm Henv}$.}
\label{fig:vp_tp}
\end{figure}

We start with the investigation on how the light curve characteristics, $V_{\rm p, 0}$ and $t_{\rm p, 0}$ defined above, are affected by $M_{\rm Henv}$. We note from \S 3.2 that an increase in the explosion energy $E$ leads to a decrease in $t_{\rm p, 0}$ (shorter duration) and a decrease of $V_{\rm p, 0}$ (brighter plateau). Therefore inferring $M_{\rm Henv}$ solely from $V_{\rm p, 0}$ (see for example \citealt{barker22,barker23}) or $t_{\rm p, 0}$ (see for example \citealt{gutierrez17a, gutierrez17b}) is not feasible without assuming a relation between explosion energy $E$ and the properties of the hydrogen-rich envelope. This assumption is not necessarily justified, as the explosion mechanism is closely related to the properties of the innermost core (\citealt{ertl16,burrows21,burrows24}), but has little to do with the outermost envelope that is decoupled from the rapid core evolution in the final years of the massive star. As shown in the upper panel of Figure~\ref{fig:vp_tp}, if $E$ is adjusted to produce the light curves with the same plateau duration $t_{\rm p, 0}$, the model with larger $M_{\rm Henv}$ is brighter. Similarly, the model with larger $M_{\rm Henv}$ will have longer plateau duration if $E$ is modified such that the light curves have the same magnitude $V_{\rm p, 0}$. This behavior suggests that, by carefully adjusting $E$ to normalize the sample of SNe II light curves to have the same $t_{\rm p,0}$, their plateau magnitudes $V_{\rm p,0}$ can serve as the measurements of the $M_{\rm Henv}$.   

Motivated by the above discussion, we investigate the relation between $V_{\rm p,0}$ and $t_{\rm p,0}$, a method frequently employed to constrain the nature of transients (\citealt{kasen09,villar17,khatami23}). The result is shown in Figure \ref{fig:vp_tp}, which reveals several distinct features: (1) models with lower $M_{\rm Henv}$ occupy the region of smaller $t_{\rm p,0}$, irrespective of the variation in $M_{\rm ZAMS}$; (2) for all the progenitor models, they follow almost the same $V_{\rm p,0}$-$t_{\rm p,0}$ scaling relation when $E$ varies, i.e., $V_{\rm p,0}\propto 11.95\,\times\,{\rm log}\,t_{\rm p,0}$ (The dotted lines).

Scaling analysis readily explains the two features: (1) From Equation \ref{Eq:scaling_this_work}, for a progenitor with given envelope properties (i.e., given the same $M_{\rm Henv}$ and $R$), the range of $E$, i.e., 0.1\,-\,2.5 foe, will create 0.24 dex difference in $t_{\rm p,0}$, smaller than the 0.40 dex difference created by the variation in $M_{\rm Henv}$, which ranges from $\sim$\,3 to 14\,$M_{\rm \odot}$. To have the same $t_{\rm p, 0}$, the model with the lowest $M_{\rm Henv}$ is required to explode with $E$ lower by 2.4 dex than that of the model with the largest $M_{\rm Henv}$, a difference much larger than the range of $E$ considered in this work ($\sim$\,1.4 dex). Hence, models with small $M_{\rm Henv}$ occupy the region of small $t_{\rm p,0}$; (2) By eliminating $E$ in Equation \ref{Eq:scaling_this_work}, we derive 
\begin{equation} \label{Eq:scaling_Vp_tp}
V_{\rm p, 0} - 11.95\,{\rm log}\,t_{\rm p,0} \sim-1.76\,{\rm log}\,R\,-\,5.61\,{\rm log}\,M_{\rm Henv},
\end{equation}
which explains the $V_{\rm p,0}$-$t_{\rm p,0}$ scaling relation if $R$ and $M_{\rm Henv}$ are fixed.

\begin{figure}[!htb]
\epsscale{1.15}
\plotone{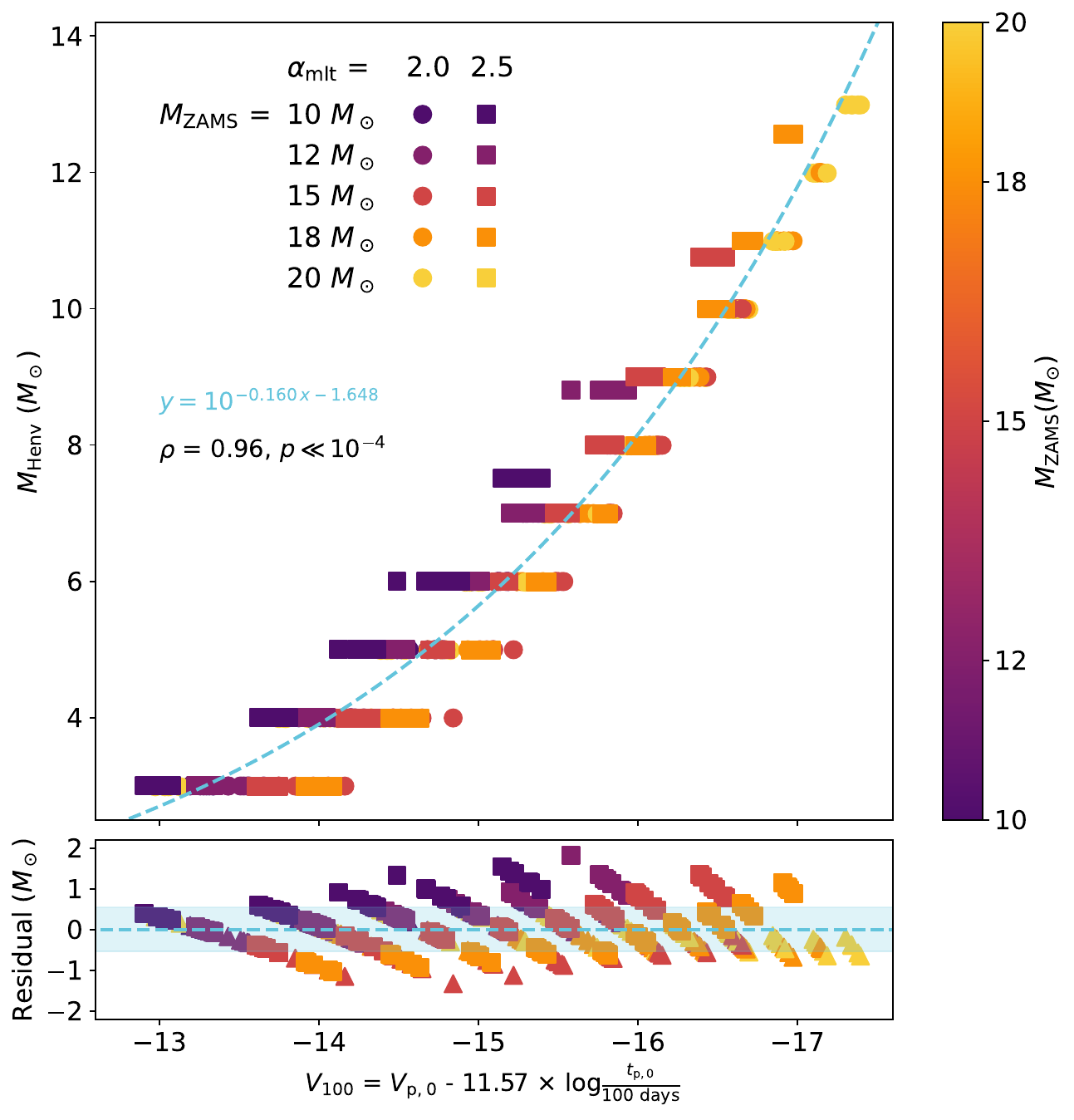}
\centering
\caption{Upper panel: The relation between $V_{\rm 100}$ and $M_{\rm Henv}$. Individual models are color coded by $M_{\rm ZAMS}$. The dashed line is the best fit. Lower panel: Deviations of each of the models from the fit. The shaded region marks the level of standard deviation.}
\label{fig:main}
\end{figure}

Motivated by Equation~\ref{Eq:scaling_Vp_tp}, we introduce a new quantity $V_{\rm 100}$ as
\begin{equation} \label{Eq:v100_definition}
    V_{\rm 100} = V_{\rm p, 0} - 11.95\times{\rm log}\frac{t_{\rm p,0}}{100~{\rm days}},
\end{equation}
i.e., the plateau magnitude $V_{\rm p,0}$ when the plateau duration is `stretched' to be 100 days by adjusting the explosion energy $E$, following Equation~\ref{Eq:scaling_Vp_tp}. It is feasible to derive this quantity observationally, as $V_{\rm p, 0}$ and $t_{\rm p, 0}$ can be determined from the observed light curves (while the effect of the $^{56}$Ni heating should be corrected for; see \S3.4). According to Equation~\ref{Eq:scaling_Vp_tp}, this quantity is determined by both the radius and the mass of the hydrogen-rich envelope, with a much stronger dependence on $M_{\rm Henv}$ than that on $R$. Figure \ref{fig:main} compares $V_{\rm 100}$ with $M_{\rm Henv}$, where a strong correlation can be immediately discerned ($\rho$\,=\,-0.96, $p\,\ll$\,0.0001). $M_{\rm Henv}$ is associated with $V_{\rm 100}$ via
\begin{equation} \label{Eq:MH_v100}
    \frac{M_{\rm Henv}}{M_{\rm \odot}} = 10^{-0.160\times V_{\rm 100} - 1.648}.
\end{equation}
The scatter in Figure \ref{fig:main} is partly attribute to the difference in $R$. The radii of the progenitor models vary from 450 to 1100\,$R_{\rm \odot}$, or 0.39\,dex, which is equivalent to a difference of 0.07\,dex in $M_{\rm Henv}$ given the same $V_{\rm 100}$, according to Equation~\ref{Eq:scaling_Vp_tp}. The 0.07\,dex difference is then transformed to a scatter of $\pm$\,0.6\,$M_{\rm \odot}$ for $M_{\rm Henv}$\,=\,7.0\,$M_{\rm \odot}$. Further, for each progenitor model, we have assumed $V_{\rm p, 0}\,\propto\,{\alpha}\,{\rm log}\,t_{\rm p,0}$ in Figure~\ref{fig:vp_tp}. The stretching factor $\alpha$ depends on both $M_{\rm Henv}$ and $R$, and ranges from 8.10 to 12.84. Fixing it to be 11.95 (Equation \ref{Eq:v100_definition}) will also contribute to the scatter. Usually $M_{\rm Henv}$ and $R$ are not determined priorly from observation, therefore these sources of scatter cannot be reduced. However, the scatter level is relatively small ($<$ 1\,$M_{\rm \odot}$ with standard deviation = 0.54\,$M_{\rm \odot}$; see the lower panel of Figure \ref{fig:main}), which in practice can be considered as the random uncertainty when applied to observation, as will be discussed in \S 4.

\subsection{Effect of the $^{56}$Ni heating}
In the previous section, we present a method to measure $M_{\rm Henv}$ for the $^{56}$Ni-free model. However, before applying these results to the observed light curves of SNe II, it is necessary to correct for the effects of the $^{56}$Ni heating. 

\begin{figure}[!htb]
\epsscale{1.15}
\plotone{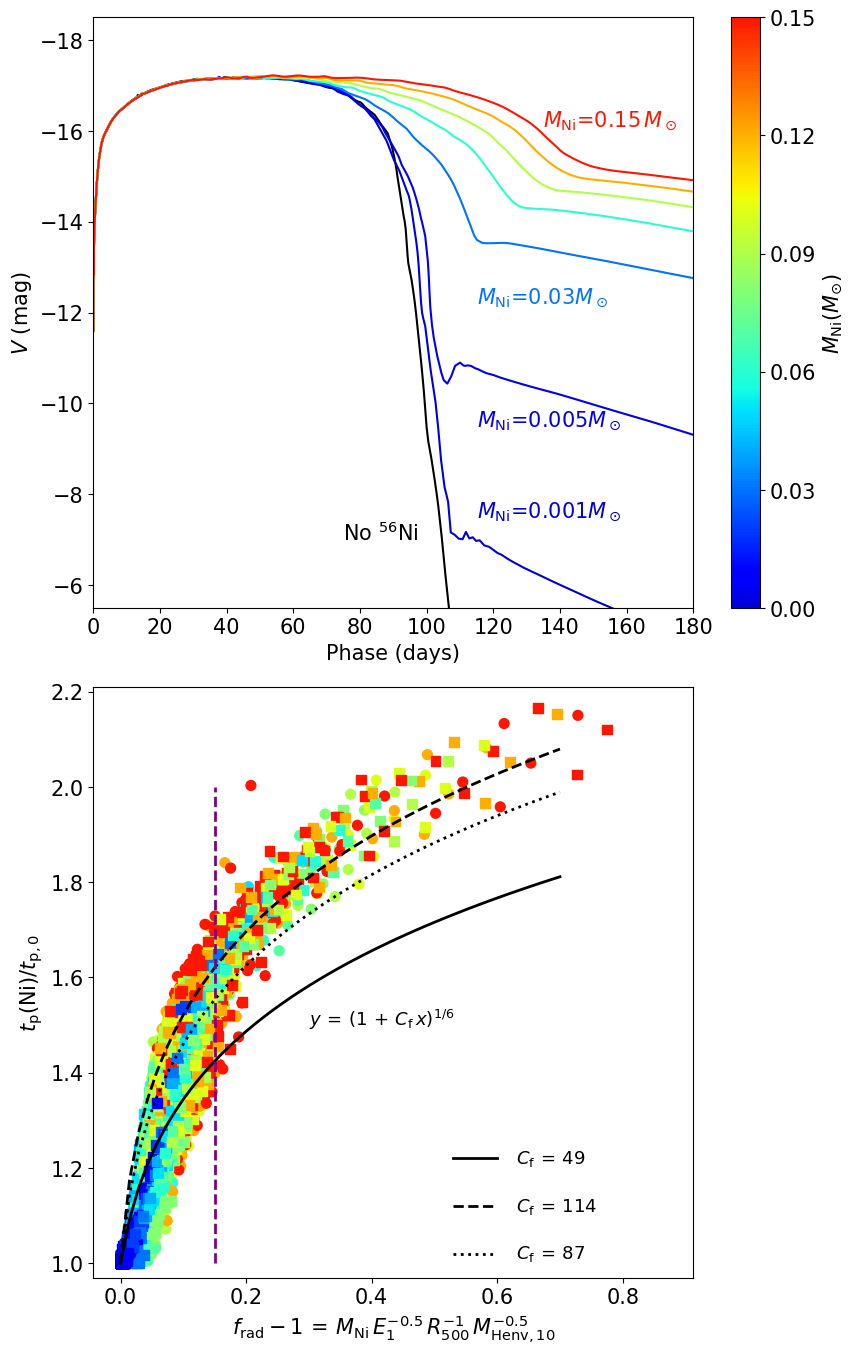}
\centering
\caption{Upper panel: The $V$-band light curve of the model with 
$\alpha_{\rm mlt}$\,=\,2.0, $M_{\rm ZAMS}$\,=\,15\,$M_{\rm \odot}$, $M_{\rm Henv}$\,=\,8.0\,$M_{\rm \odot}$, and $E$\,=\,1\,foe, but with different $M_{\rm Ni}$, as color coded by the colorbar. Lower panel: Plateau duration of models with $^{56}$Ni divided by the plateau duration for the nickle-free models, and are compared to the scaling Equation \ref{Eq:tp_ni_form2} with different values of $C_{\rm f}$. The meaning of the markers are the same as Figure~\ref{fig:scaling}. The models with $\alpha_{\rm mlt}$\,=\,2.0 evolved with standard \texttt{Dutch} wind are specially labeled by the pink stars. The purple vertical line roughly marks the upper limit of the parameter space of previous model surveys (\citealt{kepler16,goldberg19})}
\label{fig:ni_example}
\end{figure}

When the photons generated by the $^{56}$Ni/Co/Fe decays diffuse through the inner ejecta and encounter the recombination front, the propagation of the recombination wave is delayed due to the continued ionization of the hydrogen-rich envelope by these photons. Observationally, the extra heating from $^{56}$Ni extends the plateau duration $t_{\rm p}$. The effects of $^{56}$Ni on the model with $M_{\rm ZAMS}$\,=\,15\,$M_{\rm \odot}$, $\alpha_{\rm mlt}$\,=\,2.0, $M_{ \rm Henv}$\,=\,8.0\,$M_{\rm \odot}$, $E$\,=\,1.0\,foe is shown in the upper panel of Figure \ref{fig:ni_example}.

In the literature (see for example \citealt{kasen09,kepler16,goldberg19,kozyreva19}), the effects of $^{56}$Ni on SNe II light curves have been extensively studied. It has been demonstrated that the $^{56}$Ni heating has little effect on the magnitude of the plateau, while it significantly extends the plateau duration. The amount of the extension can be estimated as 
\begin{equation} \label{Eq:tp_ni_form1}
    t_{\rm p} = {t_{\rm p,0}} \times f_{\rm rad}^{1/6},
\end{equation}
where $f_{\rm rad}$ is the function of $M_{\rm Ni}$, $M_{\rm Henv}$, $R$ and $E$:
\begin{equation} \label{Eq:tp_ni_form2}
    f_{\rm rad} = 1 + C_{\rm f}\,M_{\rm Ni,1}\,E_{\rm 1}^{-1/2}\,R_{\rm 500}^{-1}\,M_{\rm Henv, 10}^{-1/2}.
\end{equation}
Here, $M_{\rm Ni,1}$, $R_{\rm 500}$, $M_{\rm Henv, 10}$ and $E_{\rm 1}$ are $M_{\rm Ni}$, $R$, $M_{\rm Henv}$ and $E$ in the units of 1\,$ M_{\rm \odot}$, 500\, $R_{\rm \odot}$, 10\,$M_{\rm \odot}$ and 1 foe, and $C_{\rm f}$ is the normalized constant that depends on the model grid (\citealt{kasen09,kepler16,goldberg19}).

In \citet{kepler16} and \citet{goldberg19}, the effect of the Ni-heating was investigated in a relatively narrow range of $f_{\rm rad}$\,=1.00\,-\,1.15, while the model grid in this work encompasses a largely expanded parameter space, and $f_{\rm rad}$ ranges from 1.00 to 1.70.
Applying Equation \ref{Eq:tp_ni_form2} to the models with $f_{\rm rad}\,<$1.15, we derive $C_{\rm f}$ = 49, as shown in the lower panel of Figure \ref{fig:ni_example}. For comparison, the result from \citet{goldberg19}, where $C_{\rm f}$\,=\,87, is also plotted. The result in this work is more consistent with that from \citet{kepler16} where $C_{\rm f}$\,=\,50. However, unlike the results in previous works, we find that a single value of $C_{\rm f}$ can not provide a reasonably good fit to the entire model grid. When $f_{\rm rad}$ increases to $>$\,1.2, we find significant deviations, where the fit to these models returns $C_{\rm f}$\,=\,114. The difference in $C_{\rm f}$ is not surprising: indeed, the models with $f_{\rm rad}>$\,1.2 are large in $M_{\rm Ni}$ but low in $E$ and $M_{\rm Henv}$. The amount of radioactive energy, which is proportional to $M_{\rm Ni}$, dominates over internal energy at recombination, which scales as $E$. The recombination is therefore more affected than in the models with $f_{\rm rad}\,\le$\,1.2; it is reflected in the increase in $C_{\rm f}$.

Similar to \citet{goldberg19}, where the scaling relation connecting $t_p$ with $M_{\rm Ni}$, $M_{\rm Henv}$, $E$, and $R$ was suggested for $^{56}$Ni-rich events, we perform a power-law fit to the models with $M_{\rm Ni}$\,$\geq$\,0.04\,$M_{\rm \odot}$, and find
\begin{equation} 
\label{Eq:tp_physics_scaling}
\begin{split}
    {\rm log}\,\frac{t_{\rm p}}{{\rm days}} = 2.35\,+\,0.21\,{\rm log}\,M_{\rm Ni,1}\,+\,0.55\,{\rm log}\,M_{\rm Henv, 10}\\-0.31\,{\rm log}\,E_{\rm 1}\,-\,0.13\,{\rm log}\,R_{\rm 500}.
\end{split}
\end{equation}
The coefficients are in good agreement with those in \citet{goldberg19}, except for the dependence on $R$, which is found to be as small as 0.02 in \citet{goldberg19}. For the model grid in this work, even for $^{56}$Ni-rich events, the dependence of plateau duration $t_{\rm p}$ on $R$ is not negligible.

Although Equations \ref{Eq:tp_ni_form2} and \ref{Eq:tp_physics_scaling} provide useful ways to estimate the effect of $^{56}$Ni on the duration of the SNe II light curve once $M_{\rm Ni}$, $E$, $M_{\rm Henv}$ and $R$ are determined, in practice we are always facing with the inverse problem, i.e., extracting these information from the light curves. It is therefore important to establish a method to estimate the effects of the $^{56}$Ni heating from observables rather than the physical properties of the explosion.
 
\begin{figure}[!htb]
\epsscale{1.15}
\plotone{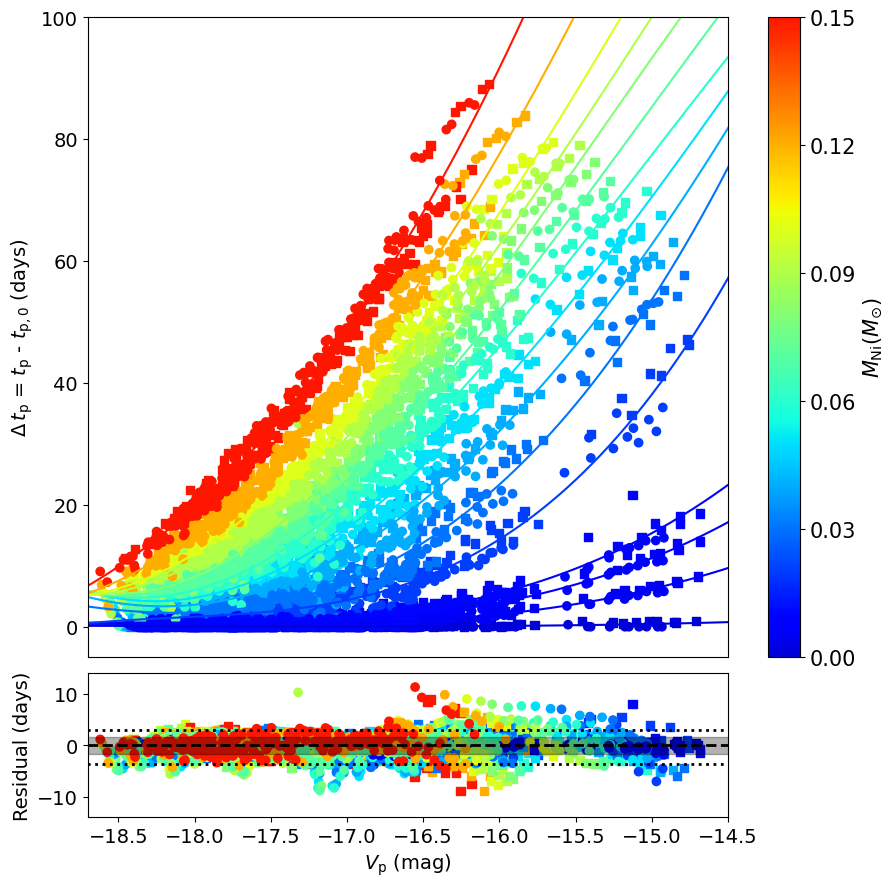}
\centering
\caption{Upper panel: The comparison between the amount of plateau extension $\Delta t_{\rm p}$ and the plateau magnitude $V_{\rm p}$. Models with different $^{56}$Ni mass are color coded by the color bar. The third-degree polynomial fits are shown by the solid lines. The meaning of the markers are the same as Figure~\ref{fig:scaling}. Lower panel: The residuals of the third-degree polynomial fits. The shaded region represents the standard deviation (1.63 days). The dotted lines represent the 95\% CI (3.35 days).}
\label{fig:vp_ni_extension}
\end{figure}

In Figure \ref{fig:vp_ni_extension}, the plateau magnitude $V_{\rm p}$ is compared with the extension of the plateau duration, $\Delta t_{\rm p}\,\equiv\,t_{\rm p}\,-\,t_{\rm p,0}$. We find an important feature, i.e., once $M_{\rm Ni}$ is fixed, the difference in the plateau duration $\Delta t_{\rm p}$ is primarily determined by $V_{\rm p}$, and bright events tend to be less extended. The relation between $\Delta t_{\rm p}$ and $V_{\rm p}$ is approximated by third-degree polynomial as 
\begin{equation}
\label{Eq:poly_fit}
\begin{split}
    \frac{\Delta t_{\rm p}}{10\,{\rm days}}
    =A\times(V_{\rm p} + 18)^3 + B\times(V_{\rm p} + 18)^2 \\+ C\times(V_{\rm p} + 18) + D,
\end{split}
\end{equation}
where the coefficients $A,B,C,D$ depend on $M_{\rm Ni}$, and are listed in Table~\ref{tab:poly_fit}. The standard deviation of the residuals is 1.63 days, as shown by the shaded region in the lower panel of Figure~\ref{fig:vp_ni_extension}, and the 95\% confidence interval (CI) is defined by $\pm$\,3.35 days. We note here that Equation~\ref{Eq:poly_fit}
is only applicable for the range of $V_{\rm p}$ shown in Figure~\ref{fig:vp_ni_extension}.

Unlike Equations \ref{Eq:tp_ni_form2} and \ref{Eq:tp_physics_scaling}, the correction of $t_{\rm p}$ for the effects of $^{56}$Ni heating, using Equation \ref{Eq:poly_fit}, is derived empirically based on observables. This correction therefore does not require any prior knowledge on the physical properties of the explosion, except for $M_{\rm Ni}$, which can be independently and robustly measured from the radioactive tail of the light curve (\citealt{woosley88,arnett89,hamuy03,spiro14,anderson19,iron_A}) or roughly estimated from the plateau magnitude (\citealt{hamuy03,kasen09,valenti16}; see also \S 4).  

\begin{deluxetable}{c|cccc}[t]
\centering
\label{tab:poly_fit}
\tablehead{
\colhead{$M_{\rm Ni}$ ($M_{\rm \odot}$)}&$A$&$B$&$C$&$D$
}
\startdata
1$\times$10$^{-3}$&0.002&-0.000&-0.007&0.011\\
5$\times$10$^{-3}$&0.021&0.004&-0.003&0.027\\
8$\times$10$^{-3}$&0.033&0.023&-0.003&0.050\\
0.01&0.043&0.032&0.005&0.069\\
0.02&0.108&0.038&0.125&0.164\\
0.03&0.059&0.365&0.084&0.233\\
0.04&-0.013&0.610&0.254&0.361\\
0.05&-0.064&0.782&0.423&0.475\\
0.06&-0.095&0.854&0.670&0.602\\
0.07&-0.095&0.847&0.911&0.740\\
0.08&-0.119&0.950&1.055&0.818\\
0.09&-0.084&0.796&1.353&1.044\\
0.10&-0.066&0.757&1.541&1.200\\
0.12&-0.018&0.673&1.850&1.509\\
0.15&0.046&0.592&2.235&1.962\\
\hline
\enddata
\caption{Polynomial coefficients of Equation \ref{Eq:poly_fit} used to correct for the effect of $^{56}$Ni heating for $V$-band light curve.}
\end{deluxetable}

\subsection{Effect of the material mixing}
Here we examine how the mixing of the material affects the properties of the light curve at the plateau phase. We select a representative progenitor model with $\alpha_{\rm mlt}$\,=\,2.0, $M_{\rm ZAMS}$\,=\,15\,$M_{\rm \odot}$, $M_{\rm Henv}$\,=\,8.0\,$M_{\rm \odot}$ and $E$\,=\,0.8\,foe, and consider 2 cases: (1) the mixing of the material by artificially changing the boxcar size in \S 2.2 to examine its effect on $t_{\rm p,0}$, without introducing $^{\rm 56}$Ni; (2) the mixing of $^{\rm 56}$Ni by artificially changing the boundary of the Ni-rich region from 0.15\,$M_{\rm eje}$ (confined in the inner most region) to 0.90\,$M_{\rm eje}$ (almost fully mixed) to examine the effect of Ni distribution on the Ni heating.

\begin{figure}[!htb]
\epsscale{1.15}
\plotone{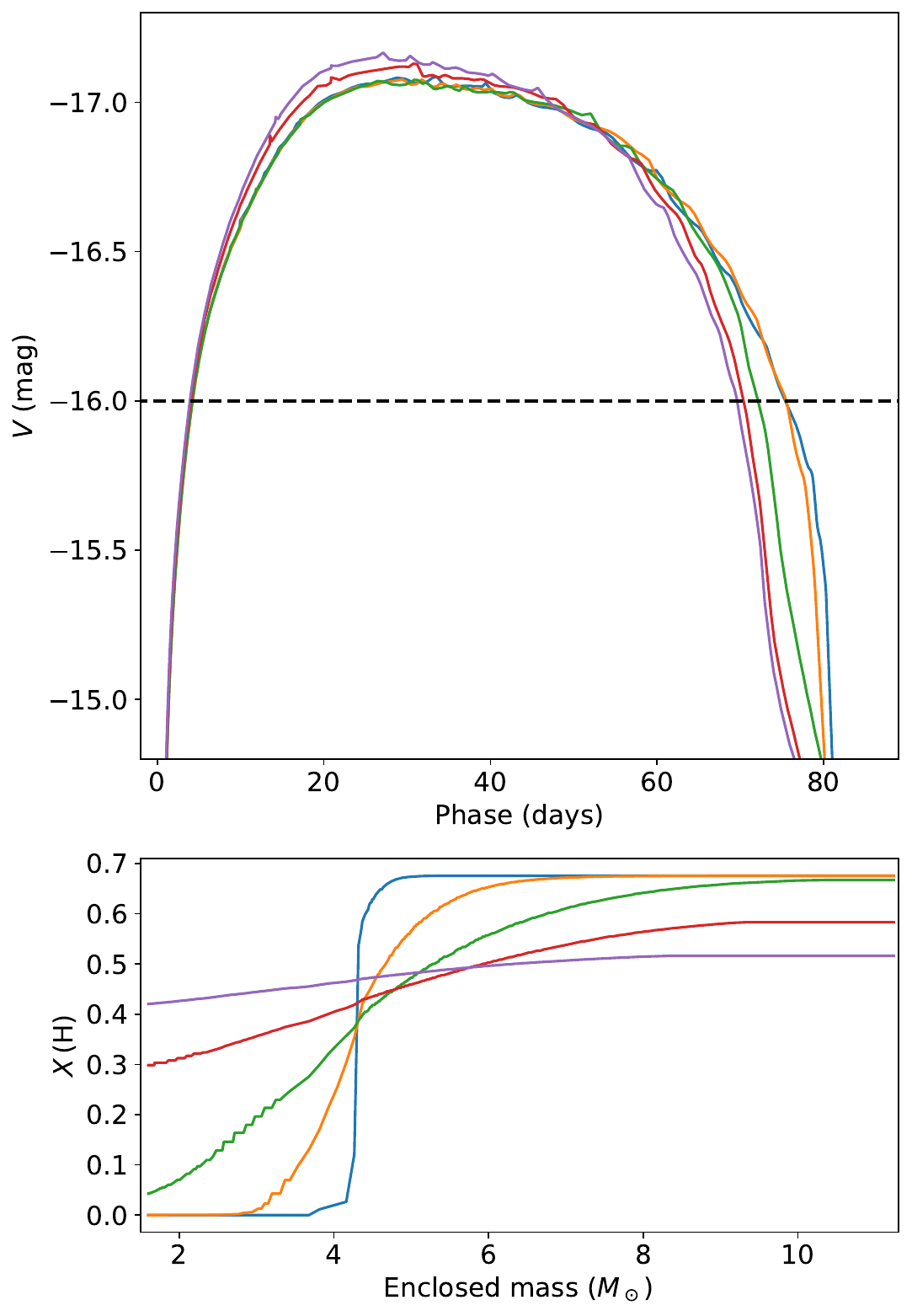}
\centering
\caption{Upper panel: The $V$-band light curve of the model with 
$\alpha_{\rm mlt}$\,=\,2.0, $M_{\rm ZAMS}$\,=\,15\,$M_{\rm \odot}$, $M_{\rm Henv}$\,=\,8.0\,$M_{\rm \odot}$, and $E$\,=\,1\,foe, but with different $M_{\rm Ni}$, as color coded by the colorbar. Lower panel: Plateau duration of models with $^{56}$Ni divided by the plateau duration for the nickle-free models, and are compared to the scaling Equation \ref{Eq:tp_ni_form2} with different values of $C_{\rm f}$. The meaning of the markers are the same as Figure~\ref{fig:scaling}. The models with $\alpha_{\rm mlt}$\,=\,2.0 evolved with standard \texttt{Dutch} wind are specially labeled by the pink stars. The purple vertical line roughly marks the upper limit of the parameter space of previous model surveys (\citealt{kepler16,goldberg19})}
\label{fig:global_mix}
\end{figure}

\begin{figure}[!htb]
\epsscale{1.15}
\plotone{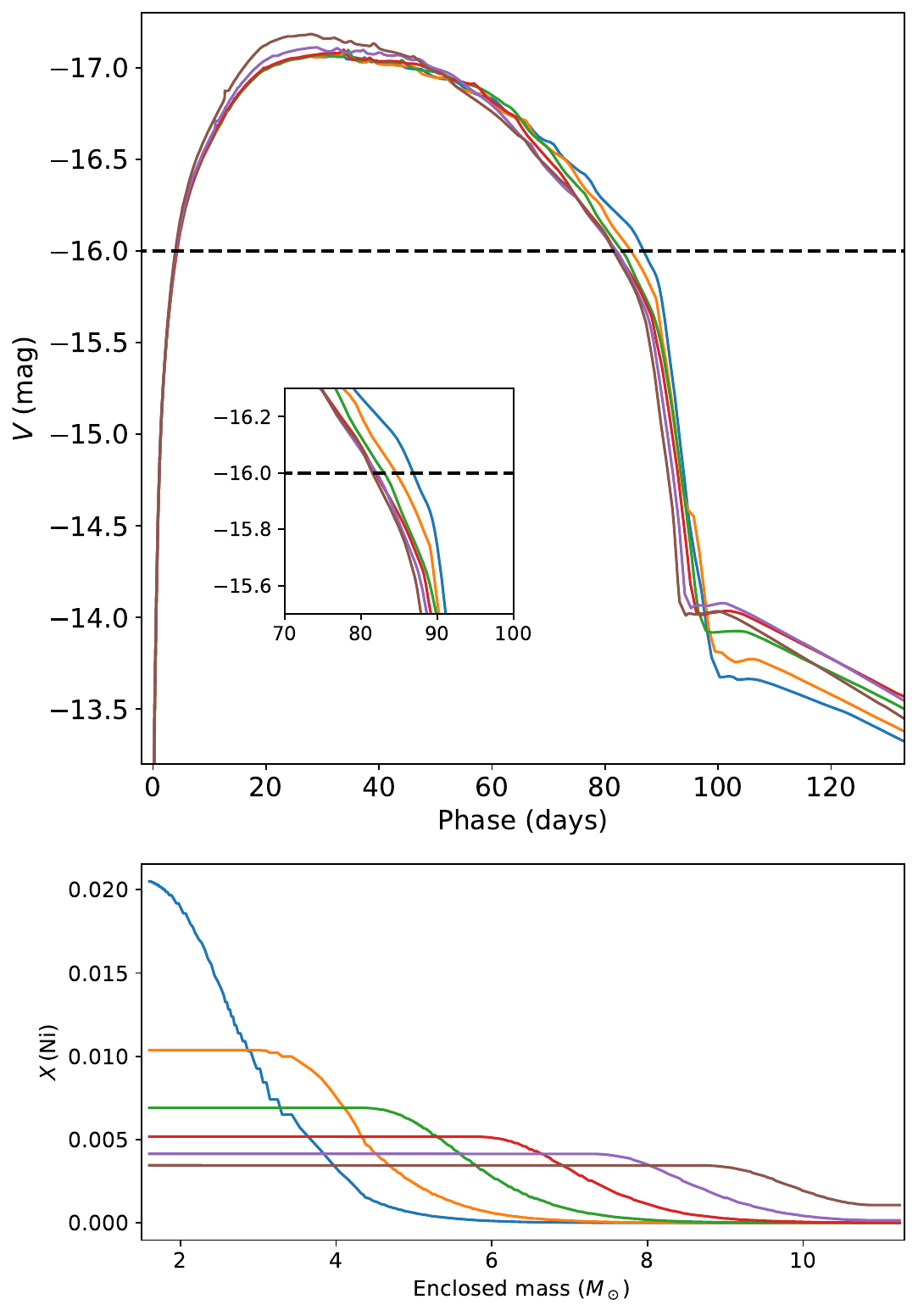}
\centering
\caption{Upper panel: The $V$-band light curve of the model with 
$\alpha_{\rm mlt}$\,=\,2.0, $M_{\rm ZAMS}$\,=\,15\,$M_{\rm \odot}$, $M_{\rm Henv}$\,=\,8.0\,$M_{\rm \odot}$, and $E$\,=\,1\,foe, but with different $M_{\rm Ni}$, as color coded by the colorbar. Lower panel: Plateau duration of models with $^{56}$Ni divided by the plateau duration for the nickle-free models, and are compared to the scaling Equation \ref{Eq:tp_ni_form2} with different values of $C_{\rm f}$. The meaning of the markers are the same as Figure~\ref{fig:scaling}. The models with $\alpha_{\rm mlt}$\,=\,2.0 evolved with standard \texttt{Dutch} wind are specially labeled by the pink stars. The purple vertical line roughly marks the upper limit of the parameter space of previous model surveys (\citealt{kepler16,goldberg19})}
\label{fig:ni_mix}
\end{figure}

\begin{itemize}
\item Global mixing.~~~The upper panel of Figure~\ref{fig:global_mix} compares the $V$-band light curves of the models with the same progenitor and explosion energy but varied boxcar sizes to mimic different degrees of large-scale material mixing. The resulting mass fractions of hydrogen are shown in the lower panel. The most significant effect is on the duration of the light curve, with the plateau becoming shorter if the hydrogen elements are mixed inward. The quantity $V_{\rm 100}$, used to estimate the hydrogen-rich envelope mass, will increase by 0.30 dex (becoming fainter). This will result in approximately a 10\% difference in $M_{\rm Henv}$ estimation according to Equation~\ref{Eq:MH_v100}.

\item $^{56}$Ni mixing.~~~The upper panel of Figure~\ref{fig:ni_mix} compares the $V$-band light curves of the models with the same progenitor and explosion energy, but with varied distributions of $^{56}$Ni. The boundaries of $^{56}$Ni range from 15\% to 90\% of $M_{\rm eje}$, representing different degrees of radioactive element mixing, from strongly confined to extensively mixed outward. The $^{56}$Ni mass is 0.03$M_{\rm \odot}$ and is assumed to be uniformly distributed within these boundaries, with slight smoothing by the default boxcar scheme. The resulting mass fractions of $^{56}$Ni are shown in the lower panel. Consistent with the findings of \citet{goldberg19}, outward mixing of $^{56}$Ni shortens the plateau duration by about 8 days compared to cases where $^{56}$Ni is confined to the inner regions. Therefore, we recommend an additional $\pm$\,4 days uncertainty when applying Equation~\ref{Eq:poly_fit} to correct for the $^{56}$Ni heating effect on plateau duration. 
\end{itemize}

In conclusion, the mixing of ordinary stellar material and $^{56}$Ni primarily affects the plateau duration. A larger degree of mixing (i.e., hydrogen mixed inward and $^{56}$Ni mixed outward) tends to shorten the plateau duration. Although the representative model in this section shows that this effect is not very large, it is important to note that our subsequent discussions are based on the default boxcar mixing scheme introduced in \S 2.2.

\section{Application to observation}
In the previous section, we have shown that the plateau phase of SNe II light curve does not provide substantial information about $M_{\rm ZAMS}$ of the progenitor, while the combination of the plateau magnitude and duration can constrain the envelope mass $M_{\rm Henv}$ within an uncertainty of 1\,$M_{\rm \odot}$. In this section, the analytical results are applied to the observed SNe II sample to establish the distribution of $M_{\rm Henv}$, which is employed to emphasize the uncertainty associated with the pre-SN mass-loss mechanism.

In this work, we collect light curve data of normal SNe II from the literature that have dense $V$-band photometric observations. The primarily sources are \citet{anderson14} and \citet{valenti16}, complemented by other well observed individual objects (Table \ref{tab:SNe sample}). The inclusion criterion is the availability of $V$-band photometry that cover both the plateau phase and the transition from the plateau to the linear decay tail, which enables the measurement of $t_{\rm p}$. The final sample consists of 100 normal SNe II.

To measure the plateau duration $t_{\rm p}$, we fit the $V$-band light curve around the drop from the plateau with Equation~\ref{Eq:valenti}, using the python routine \texttt{scipy.optimize.curve\_fit}. The main source of the uncertainty comes from the uncertainty of the explosion date, and the typical value is 5 to 10 days.
It is important to note that the measurements of $t_{\rm p}$ for the models and the observation data are different, since for the observed light curve, the maximum magnitude on the plateau is usually not well-determined. Our experiment on fitting the $^{56}$Ni-rich model light curves with Equation~\ref{Eq:valenti} reveals a systematic offset between the plateau duration measured in \S3.1 (hereafter denoted as $t_{\rm p}$), and the ones derived from Equation~\ref{Eq:valenti} fitting (hereafter denoted as $t_{\rm p,fit}$)

\begin{equation}
\label{Eq:tp_fit}
    t_{\rm p,fit} = t_{\rm p} + ~6.6~{\rm days}.
\end{equation}
For the observed SNe II, the plateau duration are corrected by Equation~\ref{Eq:tp_fit}, and the standard deviation of the residual, which is 5 days, is included in the uncertainties of the measurements. Once $t_{\rm p}$ is determined, the plateau magnitude is measured by the interpolation of the observed light curve at 0.5\,$\times\,t_{\rm p}$. The main uncertainty of $V_{\rm p}$ comes from the uncertainty of the estimations of distance and extinction.

\begin{figure}[!htb]
\epsscale{1.15}
\plotone{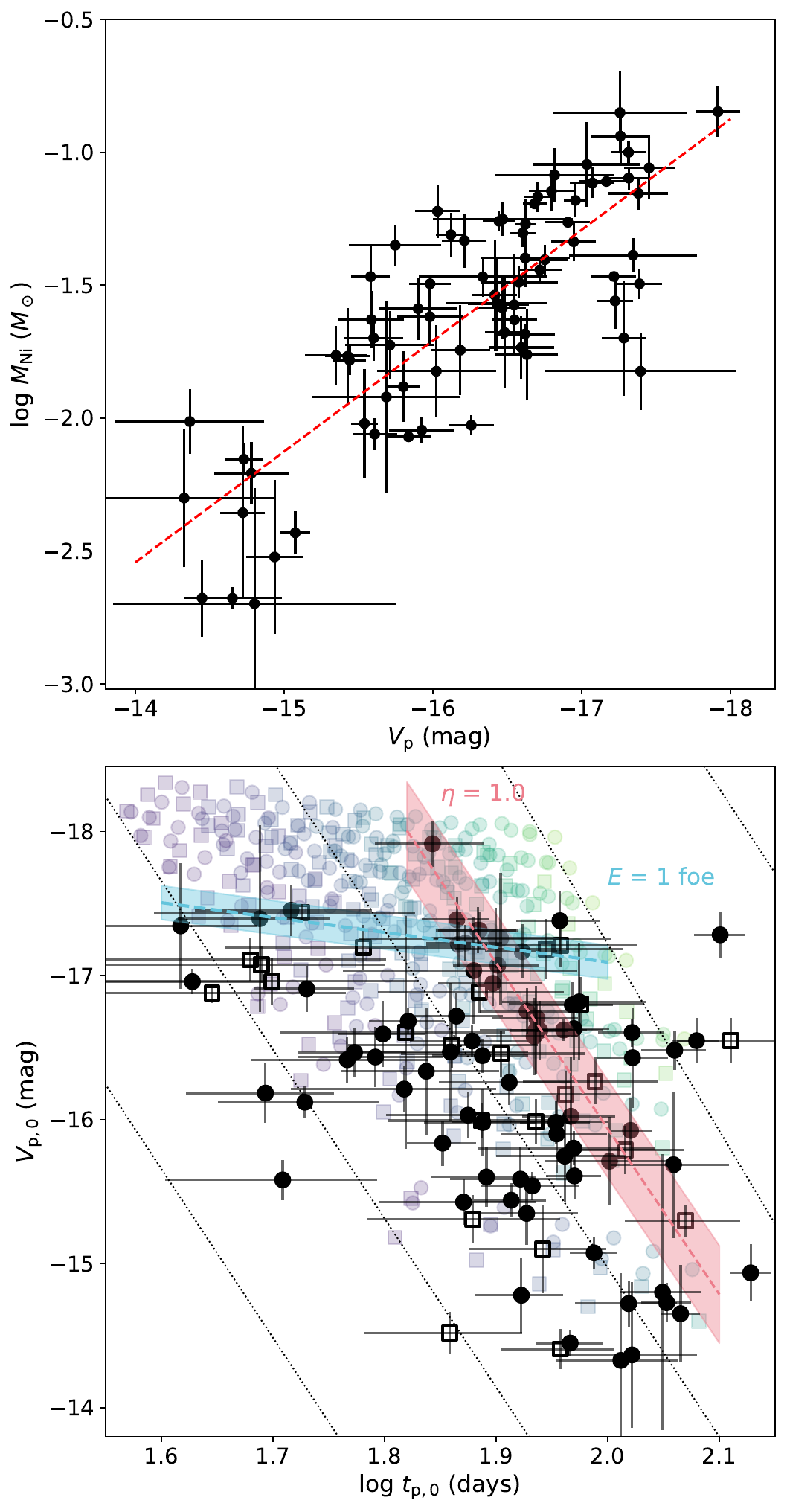}
\centering
\caption{Upper panel: The relation between plateau magnitudes $V_{\rm p}$ and the $^{56}$Ni masses $M_{\rm Ni}$ of SNe II with well constrained $M_{\rm Ni}$ from the radiative tail ($N$\,=\,79). The red dashed line is the linear regression to the data (Equation~\ref{Eq:vp_ni}). Lower panel: The comparison of the plateau magnitudes and the plateau duration, corrected for the $^{56}$Ni heating, of the SNe II sample in this work ($N$\,=\,99). Objects with and without well constrained $M_{\rm Ni}$ are labeled by the black dots and open squares respectively. The light blue strip indicates the range of models with $E$\,=\,1\,foe, while the pink strip indicates the models with wind efficiency $\eta$\,=\,1.0 (standard stellar wind). The transparent points are the models shown in the lower panel Figure~\ref{fig:vp_tp}.}
\label{fig:obs}
\end{figure}

Before measuring $t_{\rm p,0}$, it is necessary to determine $M_{\rm Ni}$ to correct for the effect of ${56}$Ni heating on the plateau duration, as discussed in \S3.4. We collect $M_{\rm Ni}$ from the literature, which is measured from the luminosity of the radioative tail. A correlation between the plateau magnitude and the $^{56}$Ni mass was firstly reported by \citet{hamuy03}, and confirmed by many subsequent works (\citealt{kasen09,valenti16}). Among the 100 SNe II in our sample, 80 of them have well constrained $M_{\rm Ni}$, and they are connected with $V_{\rm p}$ through

\begin{equation}
\label{Eq:vp_ni}
    {\rm log}\frac{M_{\rm Ni}}{M_{\rm \odot}} = -0.385\times V_{\rm p} -7.851,
\end{equation}
as shown in the upper panel of Figure~\ref{fig:obs}. The standard deviation of the residual is 0.24 dex. For objects without independent measurement on the radiative tail, their $M_{\rm Ni}$ are determined by Equation~\ref{Eq:vp_ni}. The uncertainty of $V_{\rm p}$ is propagated to that of $M_{\rm Ni}$.

After determining $t_{\rm p}$ and $M_{\rm Ni}$, the extension of plateau by the $^{56}$Ni heating, $\Delta t_{\rm p}$, is corrected through Equation~\ref{Eq:poly_fit}. We use Monte Carlo techniques to estimate the uncertainty. For each object, We perform 1000 simulations. In each trial, the uncertainty of $V_{\rm p}$ is randomly assigned assuming Gaussian distribution. Because $M_{\rm Ni}$ is estimated from the luminosity of the tail or the plateau magnitude, the uncertainties of which mainly come from the distance and extinction estimations, so we assume $\sigma$log\,$M_{\rm Ni}$\,=\,0.4\,$\times \,\sigma V_{\rm p}$ and randomly assigned to log\,$M_{\rm Ni}$. Here $\sigma$log\,$M_{\rm Ni}$ and $\sigma V_{\rm p}$ are the uncertainties of log\,$M_{\rm Ni}$ and $V_{\rm p}$. With $V_{\rm p}$ and $M_{\rm Ni}$ kept fixed, the plateau duration extended by the $^{\rm 56}$Ni heating, $\Delta t_{\rm p}$, is derived from Equation~\ref{Eq:poly_fit}. The uncertainty of $\Delta t_{\rm p}$ is the standard deviation of the 1000 measurements. The plateau duration without the $^{56}$Ni heating, $t_{\rm p,0}$, is then determined. Here we do not attempt to correct for the effect of $^{\rm 56}$Ni heating on the plateau magnitude $V_{\rm p}$ because such effect is significant only when the plateau is faint but $M_{\rm Ni}$ is large, which is not seen in observation (Equation~\ref{Eq:vp_ni}). For convenience, in the following text, we simply assume $V_{\rm p}$\,=\,$V_{\rm p,0}$.
The comparison of log\,$t_{\rm p,0}$ and $V_{\rm p,0}$ is shown in the lower panel of Figure~\ref{fig:obs}.

Two interesting features can immediately be discerned: (1) the observed SNe II cover a much broader range than that predicted by the standard stellar wind models ($\eta$\,=\,1.0 and $\alpha_{\rm mlt}$\,=\,2.0; the pink strip) in the log\,$t_{p,0}$-$V_{\rm p}$ diagram; (2) most SNe II have $E$ less than 1.0 foe (the light blue strip). The diversity in the plateau duration and magnitude of SNe II, as well as the lack of correlation, has also been reported by \citet{anderson14}, while in their work, the plateau duration was not corrected for the effect of the $^{56}$Ni heating, and the magnitude was defined at the maximum light of the initial peak. Although the uncertainty is relatively large, we find that the range of the observed SNe II in the log\,$t_{p,0}$-$V_{\rm p,0}$ diagram can be fully accounted for by the models in this work, with $M_{\rm env}$ ranging from 3 to 14\,$M_{\rm \odot}$. The mean value of $M_{\rm Henv}$ is 6.75\,$M_{\rm \odot}$, and the standard deviation is 2.98\,$M_{\rm \odot}$.

\begin{figure}[!htb]
\epsscale{1.15}
\plotone{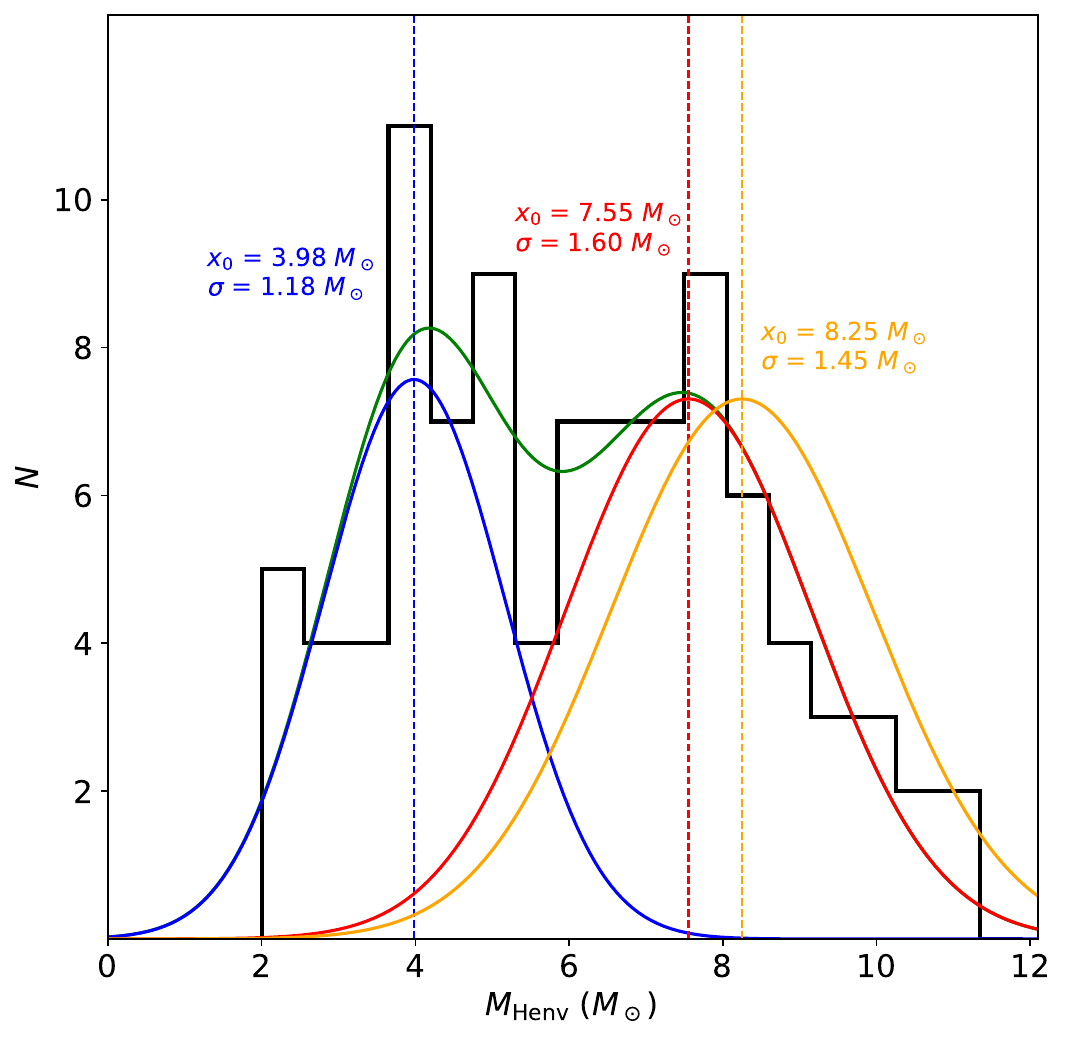}
\centering
\caption{Distribution of $M_{\rm Henv}$ derived from $V$-band light curve modeling. The distribution can be characterized by a double-Gaussian fit (green line), with the individual Gaussian components shown separately (blue and red lines). The IMF-weighted $M_{\rm Henv}$ distribution from \texttt{Kepler} progenitor models, smoothed with a Gaussian kernel ($\sigma$ = 1.0 $M_{\odot}$, reflecting the typical measurement error of $M_{\rm Henv}$), is shown by the orange line.}
\label{fig:mhenv_hist}
\end{figure}

The distribution of $M_{\rm Henv}$ is shown in Figure~\ref{fig:mhenv_hist}, ranging from approximately 2 to 12\,$M_{\rm \odot}$. Within this unexpectedly wide distribution, we find $M_{\rm Henv}$ appears to be bimodal, although its bimodality is not very pronounced. Specifically, we find a hint at the presence of two sub populations, with peaks around 7.55\,$M_{\rm \odot}$ and 3.98\,$M_{\rm \odot}$, as determined by fitting the distribution with two Gaussian functions. The center of the first peak falls within the typical range of $M_{\rm Henv}$ predicted by the stellar evolution models with the standard wind mass-loss scheme.

\begin{figure}[!htb]
\epsscale{1.0}
\plotone{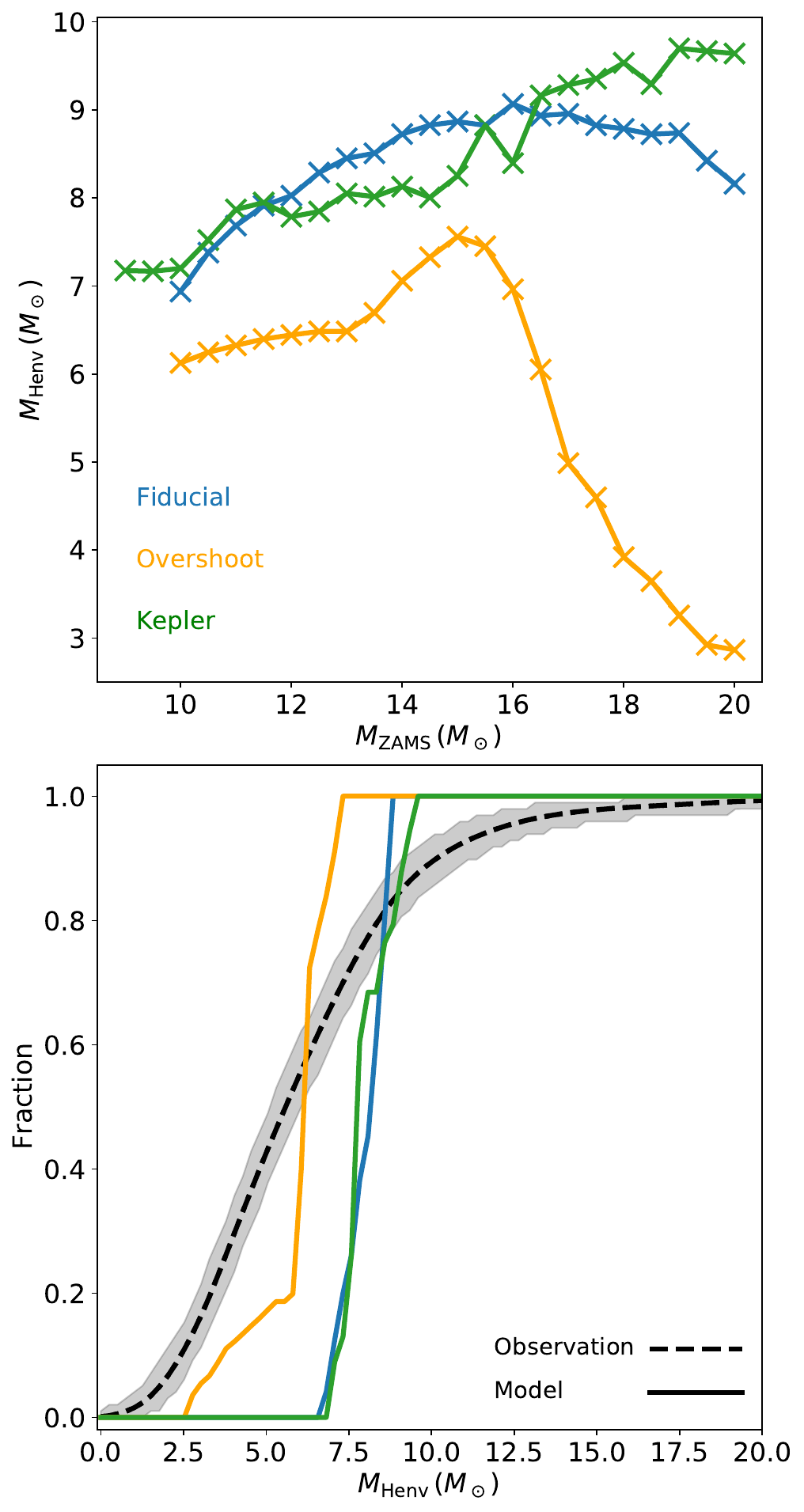}
\centering
\caption{Upper panel: The relation between $M_{\rm ZAMS}$ and $M_{\rm Henv}$ predicted by the progenitor models. Blue: fidical models; orange: strong overshoot models; green: \texttt{Kepler} models; Lower panel: The distributions of $M_{\rm Henv}$. The black dashed line is for the observed SNe II sample, and the shaded region marks the 95\% CI. The color lines are the predictions of the progenitor models.}
\label{fig:statistics}
\end{figure}

We now examine whether the distribution of $M_{\rm Henv}$ derived above, especially the possible sub-population with the smaller values of  $M_{\rm Henv}$, matches with the one expected by the mass-loss driven by stellar wind. For this purpose, we need to establish the relation between $M_{\rm ZAMS}$ and $M_{\rm Henv}$. We employ 3 progenitor model grids: (1) the fiducial model with $\alpha_{\rm mlt}$\,=\,2.0, as described in \S 2.1; (2) strong overshoot model, identical to model grid (1) except for the enhanced overshooting parameter $f_{\rm ov}$ to 0.025; (3) the grid calculated by \texttt{Kepler} (\citealt{kepler16}). For model grids (1) and (2), we assume $\eta$\,=\,1.0, and $M_{\rm ZAMS}$ ranges from 10 to 20\,$M_{\rm \odot}$ with 0.5\,$M_{\rm \odot}$ increments. The relations between $M_{\rm ZAMS}$ and $M_{\rm Henv}$ of these models are shown in the upper panel of Figure~\ref{fig:statistics}. 

For the fiducial models, the increase in $M_{\rm ZAMS}$ leads to more massive hydrogen-rich envelope if $\eta$\,=\,0.0, while at the same time, the stellar wind becomes stronger. The final $M_{\rm Henv}$ is limited to a relatively narrow range as a result of the competition between these two factors, which is similar to the $M_{\rm ZAMS}$-$M_{\rm Henv}$ relation of the \texttt{Kepler} model grid. Compared with the fiducial models, the strong overshoot models possess more massive and luminous helium cores for fixed $M_{\rm ZAMS}$, and are more efficient in the wind mass-loss. When $M_{\rm ZAMS}$ reaches to 15\,$M_{\rm \odot}$, the aforementioned balance is disrupted, and $M_{\rm Henv}$ rapidly decreases down to $\sim\,$3\,$M_{\rm \odot}$ following the continue increase of $M_{\rm ZAMS}$ to $\sim\,$20\,$M_{\rm \odot}$.

We first calculate the distributions of $M_{\rm Henv}$ expected by these theoretical models. The distribution of $M_{\rm ZAMS}$ is empirically characterized by the initial mass function. In this work, we employ the Salpeter form (\citealt{imf}) 
\begin{equation} \label{Eq:IMF}
   \frac{{\rm d}N}{{\rm d}M_{\rm ZAMS}}\,\propto\,M_{\rm ZAMS}^{-2.35}. 
\end{equation}
Using Monte Carlo techniques, a large sample ($N$\,=\,10$^4$) of progenitors with $M_{\rm ZAMS}$ ranging from 10 to 20\,$M_{\rm \odot}$ is generated, following the distribution described by Equation~\ref{Eq:IMF}. For each progenitor, we calculate its $M_{\rm Henv}$ from the $M_{\rm ZAMS}$-$M_{\rm Henv}$ relations illustrated in the upper panel of Figure~\ref{fig:statistics}. The resulting distributions of $M_{\rm Henv}$ for the different progenitor grids are shown in the lower panel of Figure~\ref{fig:statistics}. 

The most significant discrepancy between the observations and the theoretical models lies in the range of $M_{\rm Henv}$. An unusually large fraction ($\sim$\,60\%) of SNe II are found to have $M_{\rm Henv}$ lower than 6.8\,$M_{\rm \odot}$, the lower bound of the fiducial and \texttt{Kepler} models. While the strong overshoot models roughly match the lower end of the observed distribution, more than 30\% of SNe II have $M_{\rm Henv}$ exceeding the 7.2\,$M_{\rm \odot}$ upper bound predicted by these models. None of the progenitor model types can fully explain the $M_{\rm Henv}$ distribution of the SNe II sample.

However, if we consider the uncertainty in $M_{\rm Henv}$ measurement, typically around 1.2\,$M_{\rm \odot}$, and randomly assign it to the \texttt{Kepler} models, the distribution of $M_{\rm Henv}$ can be described by a Gaussian function peaking at 8.25\,$M_{\rm \odot}$, as shown in Figure~\ref{fig:mhenv_hist}. Although the central value is offset approximately by 0.7\,$M_{\rm \odot}$, considering the uncertainties in the mass-loss rate driven by RSG wind, the $M_{\rm Henv}$ distribution predicted by \texttt{Kepler} models can explain the `more-massive' sub-population in the bimodal $M_{\rm Henv}$ distribution. However, the emergence of the other (less-massive) peak requires further investigation.

The failure to reproduce the observed range of $M_{\rm Henv}$, as well as its possible bimodal distribution, prompts us to reconsider the assumptions made in this study. These assumptions primarily involve two aspects: (1) population synthesis of $M_{\rm Henv}$ and (2) light curve modeling used to constrain $M_{\rm Henv}$. These will be discussed separately in the following.

\begin{itemize}
    \item Mass-loss mechanism.~~~The population synthesis of the $M_{\rm Henv}$ distribution involves two basic assumptions:
(1) standard stellar wind ($\eta$\,=\,1.0), and (2) single-star evolution. Indeed, these two assumptions are not very solid. Regarding the assumption (1), the RSG mass-loss rate is not tightly constrained from observation. Factors such as wind clumping can enhance the mass-loss rate, while it is not included in the \texttt{Dutch} scheme of \texttt{MESA}. As demonstrated by the strong overshoot models, the change in the microphysics in the stellar-evolution calculations can also significantly affect the mass loss. For the fiducial models, we have assumed the identical convection scheme, while the convection process and overshooting can depend on $M_{\rm ZAMS}$, or vary on a case-by-case basis. The absence of a robust theory on convection contributes to the uncertainty in the mass-loss rate. Further, we have assumed non-rotating progenitor models without magnetic field, despite the significant effects these factors can have on the mass-loss rate. While these uncertainties are absorbed in the freely-adjusted $M_{\rm Henv}$ in this study, self-consistent modeling that includes all these factors is required to examine whether the $M_{\rm Henv}$ distribution of SNe II is physically plausible if the hydrogen-rich envelope is solely stripped by single star evolution. 

Aside from the wind mass-loss rate, the pre-SN mass-loss channel represents another source of uncertainty. Accumulating evidences suggest that binary interaction plays a crucial role in stripping mass from the progenitor prior to the explosion (see references in \S 1). Depending on the mass ratio of the primary/secondary star and the orbit separation, the hydrogen-rich envelope can either be fully stripped or retained. The wide range of $M_{\rm Henv}$ of SNe II can, therefore, be covered by varying orbital parameters of the binary scenario. 

\item Light curve modeling.~~~The estimation of $M_{\rm Henv}$ from observed SNe II makes use of the model grid calculated in this work based on one key assumption (among others): the progenitors of SNe II are $hydrostatic$ RSGs that have $T_{\rm eff}$ around 3200 to 3800\,K, similar to the RSGs in the Galaxy. This narrow range of $T_{\rm eff}$ constrains the relation between the ZAMS mass (or more precisely, the helium core mass) and the radius $R$ at the RSG phase. Although the pre-SN images of the progenitors of several SNe II confirm that their $T_{\rm eff}$ indeed falls within this range, several factors can change this $M_{\rm ZAMS}$-$R$ relation. 

(1) Stellar activity at the late phase. Throughout this paper, we have assumed the progenitor RSGs are in a hydrostatic state when the explosion is triggered. However, in the late phase of stellar evolution, partial ionization of hydrogen in the extensive and loosely bound envelope makes the RSG unstable against radial pulsations. These pulsations not only drive mass loss but also change the radius, therefore the RSG's radius at the time of collapse can differ from its hydrostatic state. \citet{goldberg20b} examined the effect of stellar pulsation on the resulting light curve, finding that pulsation can vary the progenitor model's radius from 760 to 1100\,$R_{\rm \odot}$, which affects the plateau luminosity by $\pm$\,0.05 dex (or 0.12 mag). However, the plateau duration is almost unaffected. According to Equation~\ref{Eq:MH_v100}, this results in a 0.02 dex (or 5\% in linear scale) difference in $M_{\rm Henv}$ estimation, which is small considering the broad range of the observed $M_{\rm Henv}$ distribution. Further, \citet{goldberg20b} used a progenitor with $M_{\rm ZAMS}$\,=\,25\,$M_{\rm \odot}$, while the $M_{\rm ZAMS}$ of SNe II progenitor are typically less massive, as indicated by both pre-SN images (\citealt{smartt15}) or late phase spectroscopy (\citealt{valenti16}), usually within the range of 10 to 15\,$M_{\rm \odot}$. Theoretical modeling suggests progenitors with $M_{\rm ZAMS}$ within this range seldom pulsate (see \citealt{yoon10_pulse} for example). From observation, \citet{soraisam18} found bright RSGs tend to have larger pulsation amplitudes, and if log\,$L$/$L_{\rm \odot}$ $\lessapprox$ 5.0, the variation in the $R$-band magnitude is $\sim\,$0.20\,mag. This will translate into 0.04 dex difference in $R$ if the pulsing RSG is still on the Hayashi line, which keeps $T_{\rm eff}$ almost constant. This variation is even smaller than the model in \citet{goldberg20b} discussed above. Given the relative low log\,$L$/$L_{\rm \odot}$ of SNe II progenitors from pre-SN images (see Figure~\ref{fig:progenitor}), their pulsations, if they occur, are expected to be weak, and this small variation in radius is not very likely to explain the diversity of SNe II light curves, and the bimodal distribution of estimated $M_{\rm Henv}$.

(2) Are the progenitors of SNe II really RSGs? In this work (and many other similar analyses), SNe II are assumed to be explosions of RSGs. Although pre-SN images have confirmed this assumption for some cases, the bimodal distribution of $M_{\rm Henv}$ suggests that some events, despite their light curves resembling those of normal SNe II, may have different origins. Recently \citet{moriya24} modeled the light curve of $blue~supergiant$ (BSG), and they find that, if the input energy $E$ and $^{56}$Ni are small, the explosions of BSGs will result in low-luminosity, short-plateau light curves. Using the method introduced in this work for such a case, the $M_{\rm Henv}$ will be estimated to be small, whereas the progenitors in their work actually have $M_{\rm Henv}$ larger than 10\,$M_{\rm \odot}$. 

In fact, during the plateau phase, the light curve is dominated by the emission from the outermost region of the envelope. From the light curve modeling, we can only infer that a strong shock wave is generated in a massive hydrogen-rich envelope. However, the exact mechanism that triggers the shock is hidden by the optically thick nature of the ejecta. For example, double detonation of white dwarfs inside a hydrogen-rich envelope (\citealt{kozyreva24}) or the collision of red giants (\citealt{dessart24b}) can also result in short and faint plateau light curves, resembling to the SNe II estimated to have low $M_{\rm Henv}$ in this work. Therefore these scenarios can potentially contribute to the peak at the low mass end in the bimodal distribution of $M_{\rm Henv}$.

\end{itemize}

The $M_{\rm Henv}$ distribution derived in this work contains rich information, and is very useful to constrain the nature of CCSNe progenitors. In \S 3.1, we have shown that the light curves of SNe II provide limited information regarding $M_{\rm ZAMS}$ of the progenitor. Measuring $M_{\rm ZAMS}$ from SNe II light curves therefore heavily relies on the correlation between $M_{\rm ZAMS}$ and $M_{\rm Henv}$, e.g., strong assumptions made on the stellar wind. This study shows that the models based on these assumptions fail to produce the observation. Several possibilities could address this discrepancy: (1) If the progenitors of SNe II resemble Galactic RSGs, then either modified mass-loss rates for single stars or binary interactions are necessary to explain the $M_{\rm Henv}$ distribution observed in SNe II; (2) Some RSG progenitors may deviate from the hydrostatic states assumed throughout this work, or some SNe II may even originate from non-RSG progenitors. However, our current study does not allow us to determine which factor is dominant, or if they all contribute equally to the observed differences between $M_{\rm Henv}$ and the predictions of single RSG star models with standard stellar winds. Further, the SNe II sample in this work is collected from the literature with various observational sources, making it difficult to estimate the possible observational biases. In the future survey, large, homogeneous sample of SNe II light curves is required to derive the representative distribution of $M_{\rm Henv}$ to better constrain the origin(s) and mechanism(s) of the mass loss of SN II progenitors.

\section{Discussion}
\subsection{Properties of bolometric light curves}
In previous sections, our discussions focused on the properties of the $V$-band light curve, as a large fraction of SNe II in our sample have only the $V$-band coverage. In this section, we provide similar analysis for pseudo bolometric light curves (i.e., computed from $UBVRI$ bands) of the same progenitor model grid. For consistency, the bolometric luminosity is transformed to bolometric magnitude via, 
\begin{equation}
    M_{\rm bol}\,=\,-2.5\,{\rm log}\frac{L_{\rm bol}}{L_{\rm bol, \odot}}\,+\,4.74,
\end{equation}
here $L_{\rm bol, \odot}$ is the solar luminosity 3.828\,$\times$\,10$^{33}$\,erg\,s$^{-1}$,
and all the measurements are done following the same method in \S 3. 

\begin{figure}[!htb]
\epsscale{1.15}
\plotone{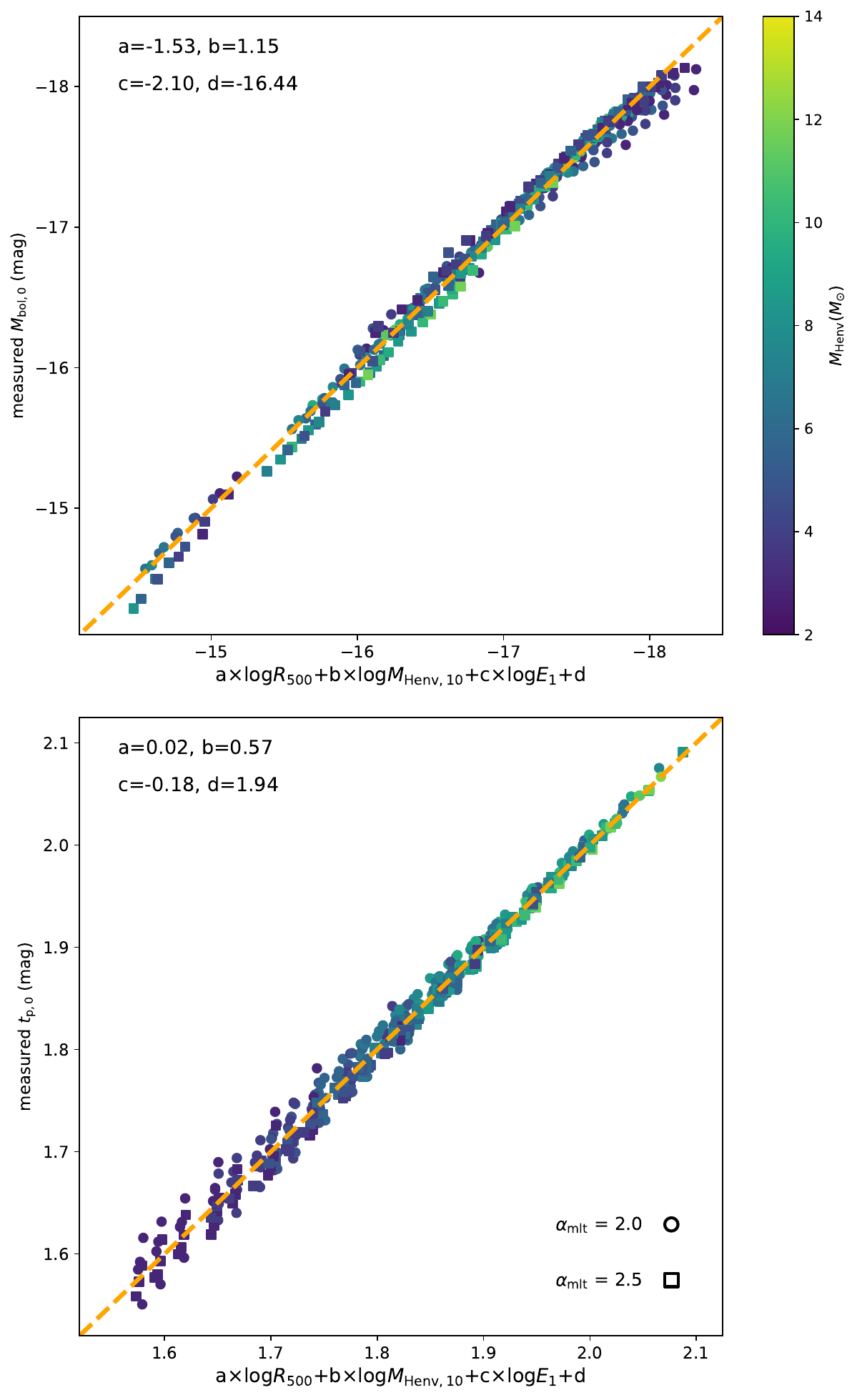}
\centering
\caption{Same as Figure~\ref{fig:scaling} but for the measurements of bolometric light curves.}
\label{fig:scaling_bol}
\end{figure}

We first derive the scaling relations for the bolometric magnitude $M_{\rm bol, 0}$ and duration of the plateau $t_{\rm p, 0}$ with $R$, $M_{\rm Henv}$ and $E$ being variables, and without the $^{56}$Ni heating:

\begin{equation} \label{Eq:scaling_this_work_bol}
\begin{split}
M_{\rm bol, 0} & \sim -1.53\,{\rm log}\,R\,+\,1.15\,{\rm log}\,M_{\rm Henv}\,-\,2.10\,{\rm log}\,E \\
{\rm log}\,t_{\rm p,0} & \sim0.02\,{\rm log}\,R\,+\,0.57\,{\rm log}\,M_{\rm Henv}\,-0.18\,{\rm log}\,E,
\end{split}
\end{equation}
and the accuracy of the fits are shown in Figure~\ref{fig:scaling_bol}. The dependence of the magnitude and duration of the bolometric light curve on the physical properties ($R$, $M_{\rm Henv}$ and $E$) are similar to $V$-band light curve, and we again confirm the weak dependence of $t_{\rm p,0}$ on $R$. Eliminating $E$ in Equations~\ref{Eq:scaling_this_work_bol}, we can similarly define 

\begin{equation} \label{Eq:m100_definition}
    M_{\rm bol, 100} = M_{\rm bol, 0} - 11.67\times{\rm log}\frac{t_{\rm p,0}}{100~{\rm days}}.
\end{equation}

\begin{figure}[!htb]
\epsscale{1.15}
\plotone{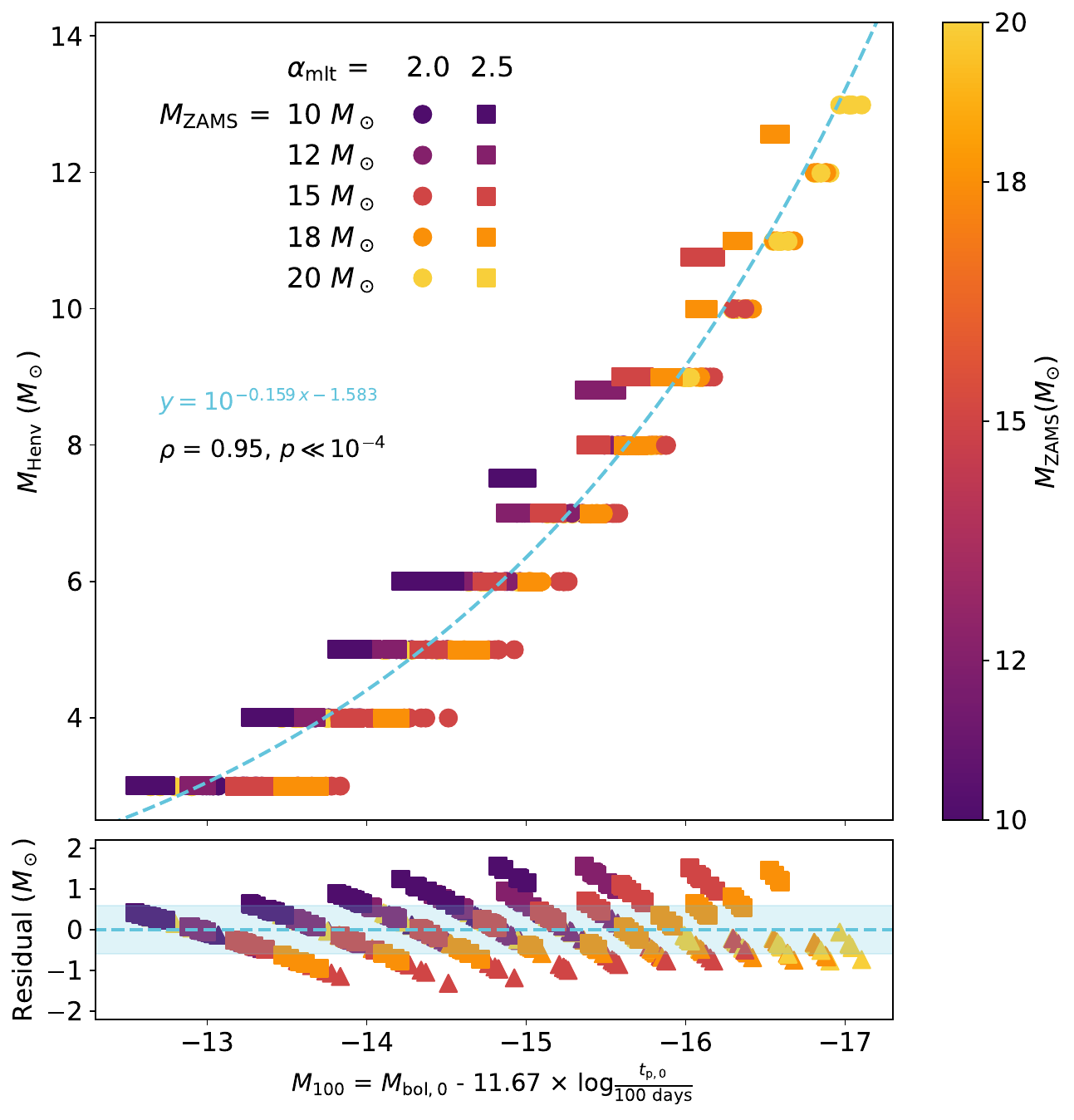}
\centering
\caption{Same as Figure~\ref{fig:main} but for the measurements of bolometric light curves.}
\label{fig:main_bol}
\end{figure}

This newly defined $M_{\rm bol, 100}$ is further compared with $M_{\rm Henv}$ in Figure~\ref{fig:main_bol}, and we derive the estimations of $M_{\rm Henv}$ based on bolometric light curves:

\begin{equation} \label{Eq:MH_v100_bol}
    \frac{M_{\rm Henv}}{M_{\rm \odot}} = 10^{-0.159\times M_{\rm 100} - 1.583}.
\end{equation}

\begin{figure}[!htb]
\epsscale{1.15}
\plotone{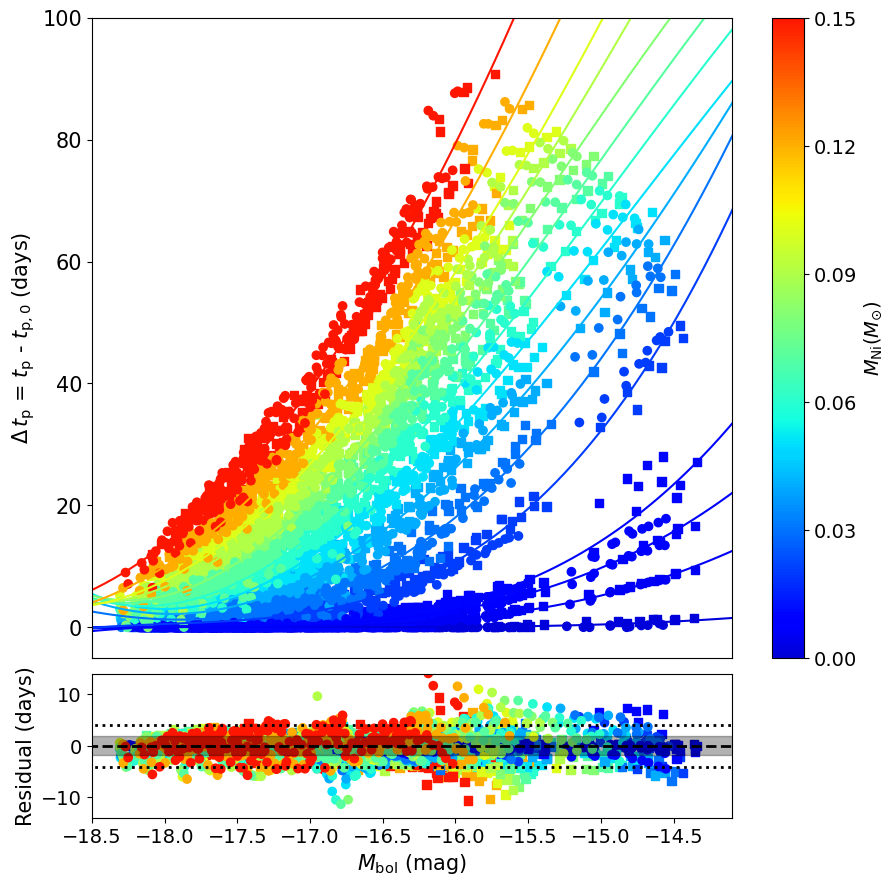}
\centering
\caption{Same as Figure~\ref{fig:vp_ni_extension} but for the measurements of bolometric light curves.}
\label{fig:vp_ni_extension_bol}
\end{figure}

For the $^{56}$Ni heating effects on the bolometric light curve, we perform similar analysis as \S3.4, and we confirm that the extension of light curve duration is solely dependent on the plateau magnitude once $M_{\rm Ni}$ is fixed, as shown in Figure~\ref{fig:vp_ni_extension_bol}, and can be described by third-degree polynomial as Equation~\ref{Eq:poly_fit}. The coefficients are summarized in Table~\ref{tab:poly_fit_bol}.

\begin{deluxetable}{c|cccc}[t]
\centering
\label{tab:poly_fit_bol}
\tablehead{
\colhead{$M_{\rm Ni}$ ($M_{\rm \odot}$)}&$A$&$B$&$C$&$D$
}
\startdata
1$\times 10^{-3}$&0.007&-0.024&0.022&0.000\\
5$\times 10^{-3}$&0.029&-0.035&0.016&0.006\\
8$\times 10^{-3}$&0.042&-0.020&0.005&0.014\\
0.01&0.076&-0.097&0.075&0.008\\
0.02&0.112&-0.015&0.095&0.053\\
0.03&0.043&0.372&-0.082&0.130\\
0.04&-0.034&0.699&-0.066&0.219\\
0.05&-0.096&0.944&0.002&0.299\\
0.06&-0.102&0.948&0.274&0.367\\
0.07&-0.110&0.972&0.479&0.458\\
0.08&-0.136&1.093&0.572&0.516\\
0.09&-0.071&0.832&0.975&0.659\\
0.10&-0.047&0.759&1.199&0.764\\
0.12&-0.008&0.677&1.524&0.996\\
0.15&0.014&0.698&1.841&1.358\\
\hline
\enddata
\caption{Polynomial coefficients of Equation \ref{Eq:poly_fit} used to correct for the effect of the $^{56}$Ni heating for the bolometric light curve.}
\end{deluxetable}

\begin{figure}[!htb]
\epsscale{1.15}
\plotone{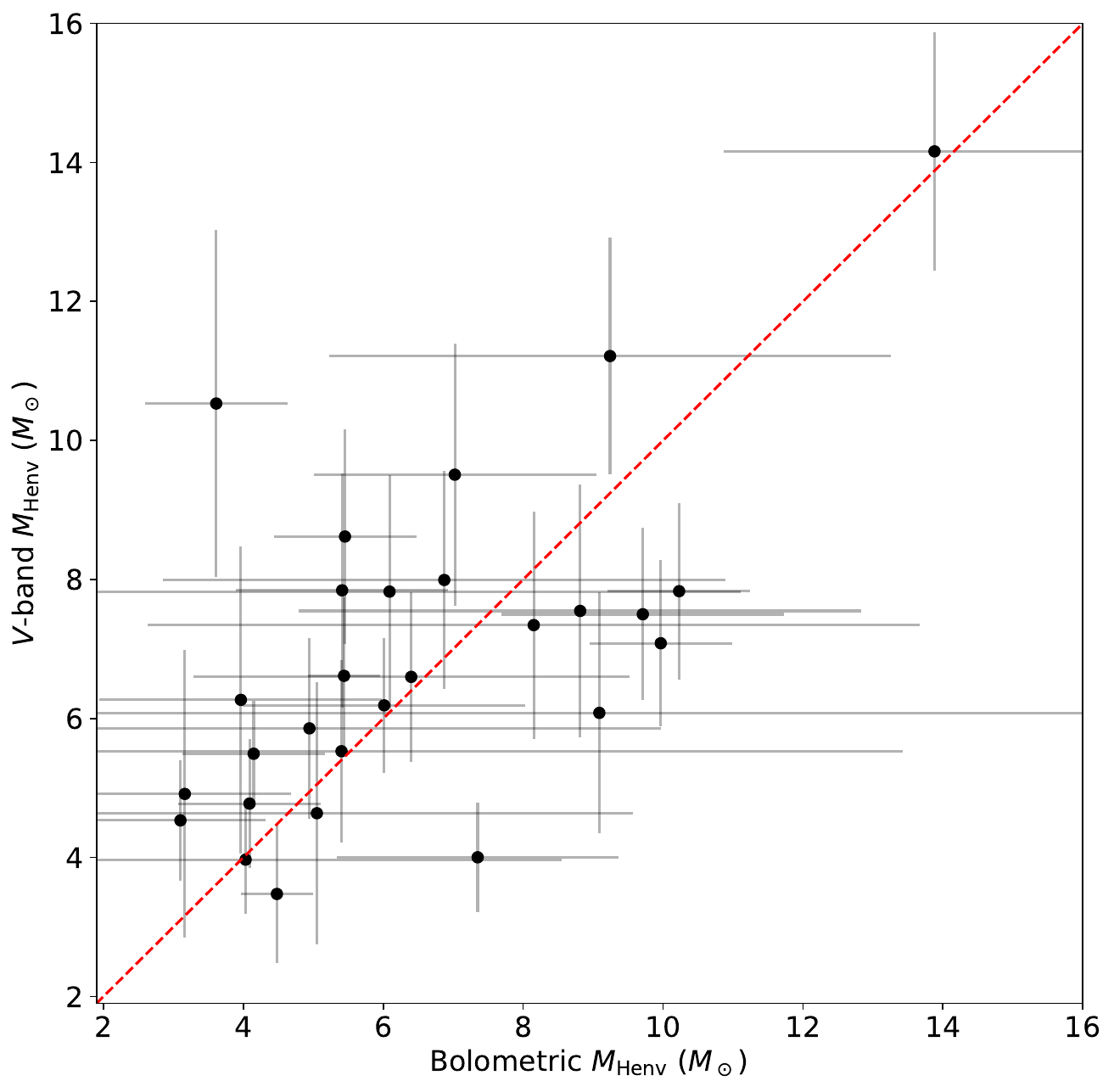}
\centering
\caption{The comparison of $M_{\rm Henv}$ measured from bolometric and $V$-band light curves, for the SNe II sample from \citet{valenti16}. The red dashed line is one-to-one correspondence.}
\label{fig:v_bolo_compare}
\end{figure}

Finally, we compare the $M_{\rm Henv}$ measured from $V$-band and bolometric light curves for SNe II from \citealt{valenti16}, as shown in Figure~\ref{fig:v_bolo_compare}. We find that the results are in general consistent within the relatively large uncertainty, and in about 70\% cases, the $M_{\rm Henv}$ estimated from bolometric light curves are smaller than those from $V$-band light curves, making the conflict between the observed $M_{\rm Henv}$ distribution and the prediction from single stellar evolution even more severe (\S 4).

\subsection{The moment when the explosion is launched}
\begin{figure}[!htb]
\epsscale{0.95}
\plotone{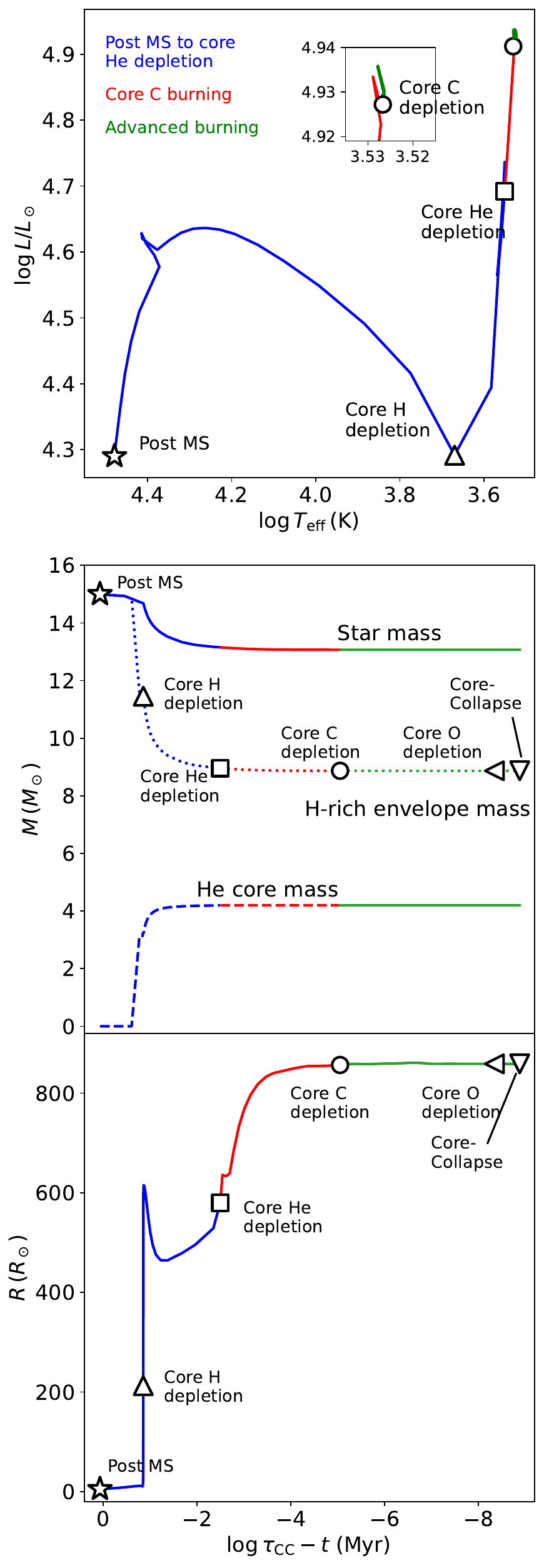}
\centering
\caption{Upper panel: The evolution track of the progenitor model with $M_{\rm ZAMS}$\,=\,15\,$M_{\rm \rm \odot}$ and $\eta$\,=\,1.0 on the HR diagram, from post MS to core-collapse. Different evolution phases are labeled by different colors. Some special checkpoints are also marked; Middle panel: The evolution of star mass (solid), hydrogen-rich envelope mass (dotted) and He core mass (dashed), shown for the time relative to the moment of core-collapse ($\tau_{\rm cc}$); Lower panel: The evolution of stellar radius.}
\label{fig:evolution_track}
\end{figure}
In this work, the progenitors are evolved to the moment when carbon in the core is exhausted, and the energy is subsequently deposited to trigger the explosion. This simplification allows us to calculate the light curves of progenitor models with $M_{\rm ZAMS}$ down to 10\,$M_{\rm \odot}$, which, in our own experiments with \texttt{MESA}, develop strong shell-burning and off-center flames, and hardly progress to the core collapse. Our goal here is to investigate whether this simplification would affect the properties of the light curve.

The upper panel of Figure~\ref{fig:evolution_track} shows the evolutionary track of the progenitor model with $M_{\rm ZAMS}$\,=\,15\,$M_{\rm \rm \odot}$ and $\eta$\,=\,1.0 on the Hertzsprung-Russel diagram (HRD). As illustrated in the middle panel of Figure~\ref{fig:evolution_track}, stripping of the hydrogen-rich envelope mainly takes place before the carbon burning phase. About 10$^4$ years after the ignition of carbon, the core starts to collapse. The mass-loss rate during this period is about 10$^{-6}$\,$M_{\rm \odot}$\,yr$^{-1}$, therefore $M_{\rm Henv}$ only changes by about 0.01\,$M_{\rm \odot}$, which is negligible in practice. The evolution of stellar radius is more complicated. In response to core hydrogen burning, the star expands, ejecting the loosely-bounded hydrogen-rich envelope. The radius reaches its local maximum ($\sim$\,600\,$R_{\rm \odot}$) when hydrogen in the core is exhausted, and shrinks again during core helium burning phase. The star expands again following core helium depletion, and settles at the (almost) constant radius until core collapse.

\begin{figure}[!htb]
\epsscale{1.0}
\plotone{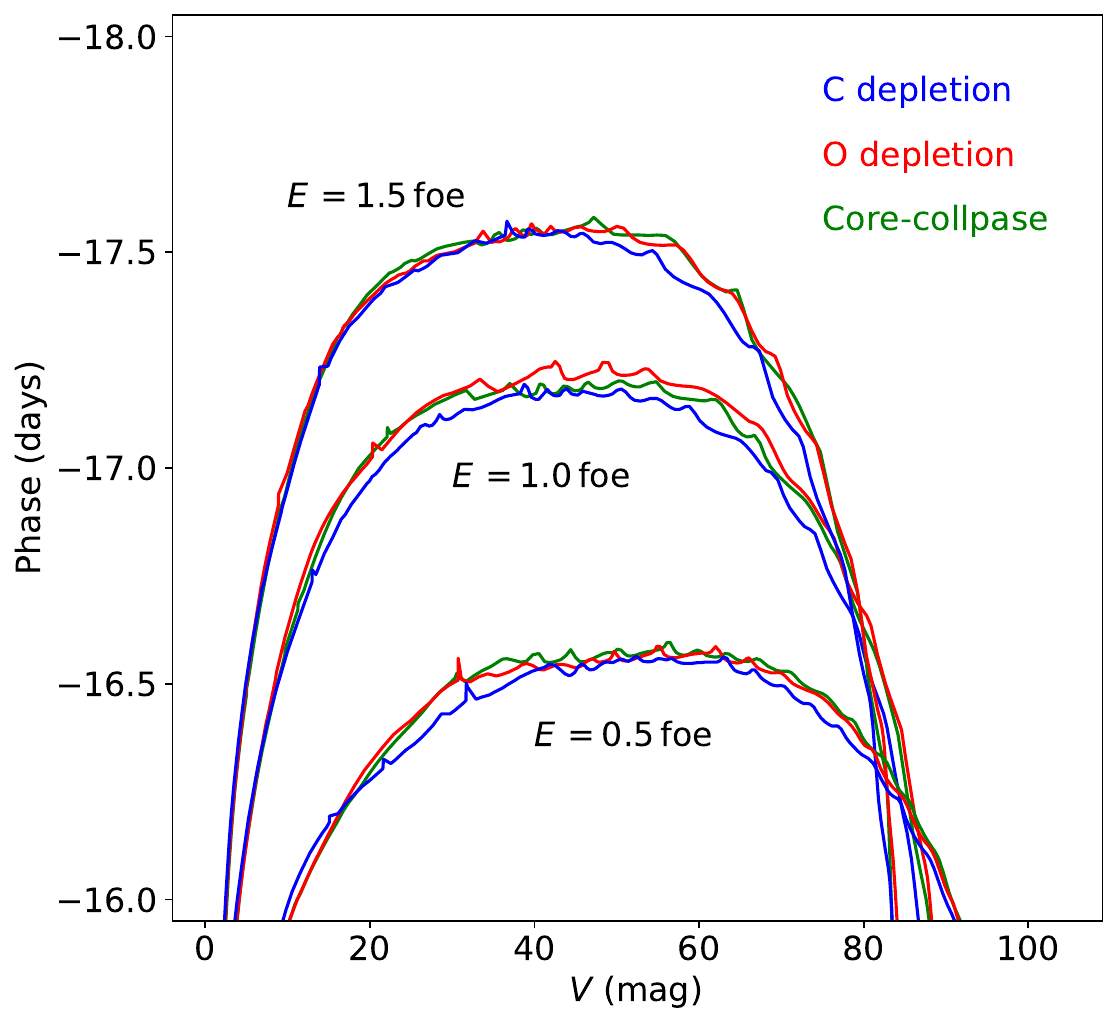}
\centering
\caption{The $V$-band light curves calculated from the progenitor structures taken at different moments: core carbon depletion (blue; this work), core oxygen depletion (red) and core-collapse (green).}
\label{fig:phase_LC}
\end{figure}

We employ two additional stellar structures, one taken at the core oxygen depletion and the other at the moment of the core-collapse, as the inputs of \texttt{STELLA}. The explosions are triggered following the procedure described in \S 2.2, and the resulting $V$-band light curves are compared with the fiducial models in this work, as shown in Figure~\ref{fig:phase_LC}. We find that the light curves are almost identical once $E$ is fixed. This is not surprising, considering that  the two main physical parameters governing the light curve properties, i.e., $M_{\rm Henv}$ and $R$, hardly evolve after the core carbon depletion.

\subsection{Velocity as an independent constraint}
In this work, we have developed a method to constrain the hydrogen-rich envelope mass of the progenitor from the light curve of SNe II. One of the main advantages of this technique is its reliance solely on photometry, and does not require information from the spectroscopy that is not always available especially for faint events. In this section, we briefly discuss whether the photospheric velocity, inferred from the minimum of the absorption features emerged in the plateau phase spectra, can provide additional constraints on $M_{\rm ZAMS}$ of the progenitor. 

As discussed in the previous sections, the properties of the light curve at the plateau phase are primarily determined by the hydrogen-rich envelope. Given the same explosion energy, models with different $M_{\rm ZAMS}$ but the same $M_{\rm Henv}$ will generate very similar light curves. Although in this work, $M_{\rm Henv}$ is arbitrarily adjusted to mimic the diverse mass-loss channels (see discussion in \S 4), the mass of the helium core $M_{\rm He,core}$ is insensitive to the remaining envelope and is almost uniquely determined by $M_{\rm ZAMS}$. Models with the same $M_{\rm Henv}$, being indistinguishable from the light curve, can be diverse in the ejecta mass $M_{\rm eje}$. Such difference is expected to manifest itself in the photospheric velocity $v_{\rm ph}$, which is associated with the explosion energy via
\begin{equation}
    E\,\sim\,\frac{1}{2}M_{\rm eje}v_{\rm ph}^{2},
\end{equation}

\begin{figure}[!htb]
\epsscale{1.0}
\plotone{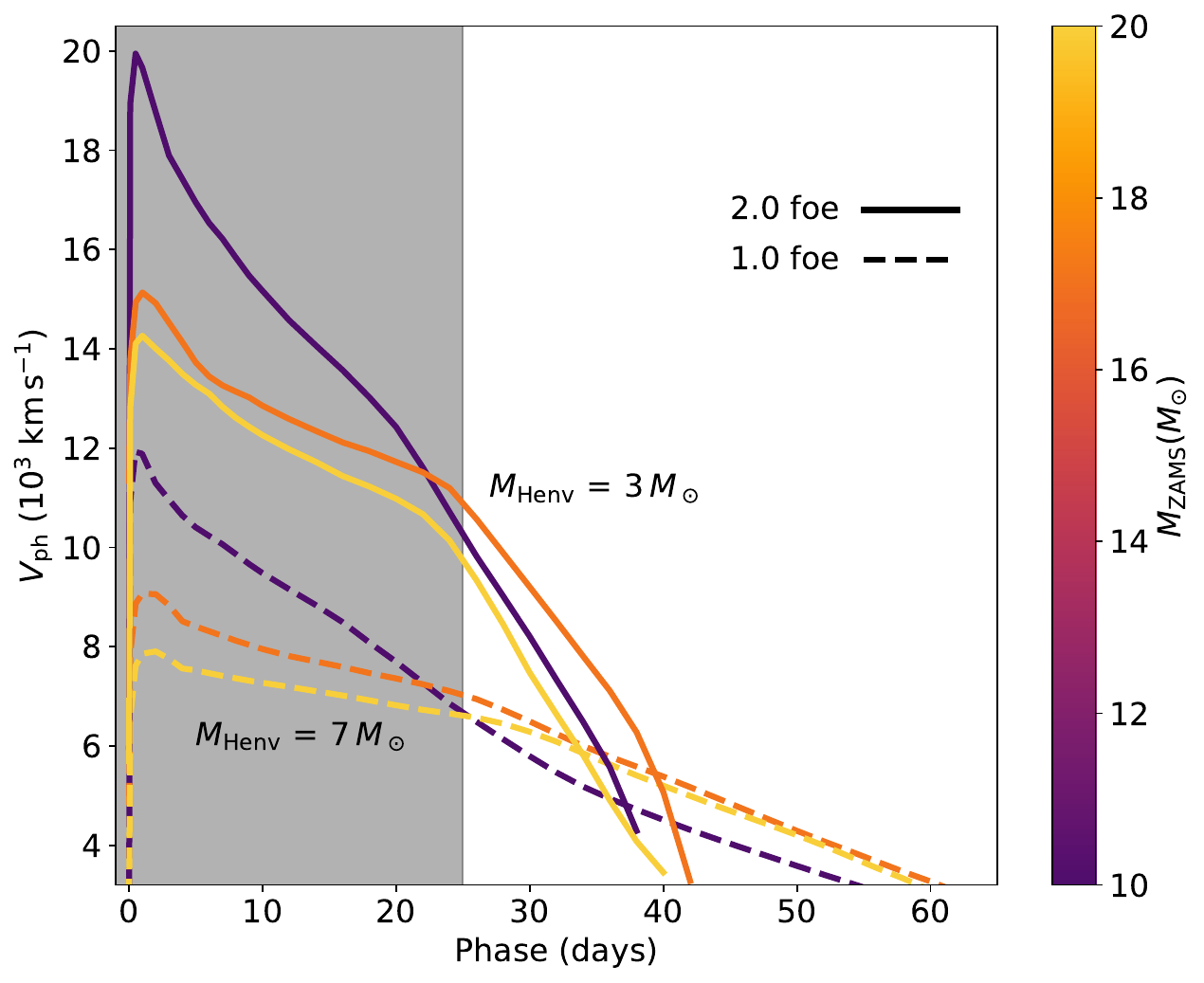}
\centering
\caption{The photosphere evolution of some typical models. For models with the same $M_{\rm Henv}$ but varied $M_{\rm ZAMS}$, the most pronounced different in $V_{\rm ph}$ arise before $\sim$\,25 days (the shaded region).}
\label{fig:vph_example}
\end{figure}

The evolution of photospheric velocities for some typical models are shown in Figure~\ref{fig:vph_example}. Here, the photosphere is defined by the point where opacity $\tau$\,=\,2/3. We select two progenitor sets, one with $M_{\rm Henv}$\,=\,7.0\,$M_{\rm \odot}$, the average value of the observed SNe II sample (see \S 4), and the other with $M_{\rm Henv}$\,=\,3.0\,$M_{\rm \odot}$. For the latter case, the variation in $M_{\rm eje}$ is the most pronounced, ranging from 4.4\,$M_{\rm \odot}$ to 7.5\,$M_{\rm \odot}$. Consistent with the findings of \citet{goldberg19}, the difference in $M_{\rm eje}$ is reflected in $v_{\rm ph}$ before 20\,days after the shock breakout, despite these models having the same $M_{\rm Henv}$. However, it is important to note that the photospheric velocity measured at the early phase is highly sensitive to the outermost density structure of the hydrogen-rich envelope, and can be significantly affected by the presence of CSM (see for example Figure 2 of \citealt{moriya23}), which is not included in the current model grid. We defer the detailed investigation on the effects of CSM on both the photospheric velocity and the light curve to future work.

At $\sim$\,30\,days after the explosion, the photosphere cools down to $\sim$\,6000 to \,7000\,K, which is set by the temperature when the recombination of hydrogen occurs. The recession of the photosphere slows down following the development of the hydrogen recombination, and the models with the same $M_{\rm Henv}$ settle down at the similar $v_{\rm ph}$. Although there are still some variations, not much can be said as these variations are relatively small and are not monotonic functions of $M_{\rm ZAMS}$.  

We now seek for the scaling relations between $v_{\rm ph}$ and other observables or physical properties. The correlation between the photospheric velocity and the luminosity of the light curve, measured at $\sim$\,50\,days after the explosion, is firstly discovered by \citet{hamuy03}, based on a sample of nearby SNe II. The physics of this correlation is then explained by \citet{kasen09}. The luminosity, assuming black-body radiation, can be expressed as
\begin{equation}
    L\,\approx\,4\pi\sigma R^2_{\rm ph} T^4_{\rm ph}.
\end{equation}
At the plateau phase, the dynamics of the ejecta can be well characterized by the homologous expansion, i.e., $v$\,($R,t$)\,=\,$R$/$t$ (\citealt{goldberg19}), and the temperature of the photosphere remains relatively constant at $T_{\rm ph}$\,$\approx$\,6000\,K, set by the hydrogen recombination, although observations indicates certain degree of variation (\citealt{valenti16}). At the given phase $t$, say, 50 days after the explosion, it is thus expected that the photospheric velocity is correlated with luminosity through $L_{\rm 50}$\,\,$\propto$\,$v_{\rm ph,50}^2$. For the model grid in this work, it is difficult to determine the phase at which the photospheric velocity should be measured. Some light curves in this work have plateau duration shorter than 40\,days. Similar to the plateau magnitude, we measure the photospheric velocity $v_{\rm ph, 0}$ at 0.5\,$\times$\,$t_{\rm p,0}$ for the 636 $^{56}$Ni-free models in the grid, and we find 
\begin{equation}\label{Eq:Vp_vph_tp}
\begin{split}
    V_{\rm p,0}\,=\,-4.84\,{\rm log}\,\frac{v_{\rm ph, 0}}{10^3\;{\rm km\,s^{-1}}}\,\\-3.75\,{\rm log}\,\frac{t_{\rm p, 0}}{100\,{\rm days}}\,-\,14.17.
\end{split}
\end{equation}
The standard deviation of the residual to the fit is 0.07\,mag. If the photospheric velocities are all measured at the similar phase, we derive $V_{\rm p,0}\,\propto \,-4.84\,{\rm log}\,v_{\rm ph, 0}$, or $L_{\rm p,0}$\,\,$\propto$\,$v_{\rm ph,0}^{1.94}$, which is in good agreement with the above analysis. The photospheric velocity is connected to the physical properties via
\begin{equation}\label{Eq:vph_scaling}
    {\rm log}\,v_{\rm ph, 0}\,\sim\,0.11\,{\rm log}\,R\;-\;0.58\,{\rm log}\,M_{\rm Henv}\;+\;0.55\,{\rm log}\,E,
\end{equation}
which confirms the degeneracy proposed by \citet{goldberg19} and \citet{goldberg20}: Equation~\ref{Eq:vph_scaling} is essentially a linear combination of Equation~\ref{Eq:scaling_this_work} through Equation~\ref{Eq:Vp_vph_tp}, and it does not contain any additional information regarding $R$, $M_{\rm Henv}$ and $E$.

\section{Conclusion}
In this work, we investigate the $V$-band light curve characteristics of SNe II, using a grid of progenitor models with various ZAMS masses, hydrogen-rich envelope masses, $^{56}$Ni masses and explosion energies calculated by \texttt{MESA}\,+\,\texttt{STELLA}. The mixing length are tuned such that the RSG progenitors have $T_{\rm eff}$ ranges from 3200 to 3800 K, similar to the RSGs observered in the Galaxy and the estimations from pre-SN images. To account for the uncertainties in the pre-SN mass-loss channels and mass-loss rates, the hydrogen-rich envelope is manually removed at the moment of the core helium depletion. We find that for these Galactic-like RSG models, the same envelope mass and explosion energy will give similar light curves, even though their $M_{\rm ZAMS}$ are different. Inferring $M_{\rm ZAMS}$ from the light curve modeling therefore can be very uncertain unless the mass-loss history is known a priori. This degeneracy, originally proposed by \citet{dessart19}, is extended in this work to encompass the typical range of the envelope masses of SNe II. 

Additionally, we establish the scaling relation between the light curve characteristics and the envelope mass $M_{\rm Henv}$, radius $R$ and explosion energy $E$ for the $^{56}$Ni-free models. We find a scaling relation for the plateau magnitude $V_{\rm p,0}$ that is very similar with previous studies (see \citealt{popov93,kasen09,kepler16,goldberg19} for examples). However,  the dependence of plateau duration $t_{\rm p, 0}$ on $R$ is surprisingly weak, as shown in our Figure~\ref{fig:scaling}, contrary to the previously proposed scaling $t_{\rm p, 0}\propto R^{1/6}$. 
Based on these equations, we develop a method to measure $M_{\rm Henv}$ by combining $V_{\rm p, 0}$ and $t_{\rm p, 0}$. We find $M_{\rm Henv}$ can be well constrained within an uncertainty of 1\,$M_{\rm \odot}$ (Figure~\ref{fig:main}). The effects of the $^{56}$Ni heating, known to potentially extend the plateau duration, are also thoroughly discussed in this study. We find that once the mass of $^{56}$Ni is fixed, the amount of plateau extension is almost uniquely determined by the plateau magnitude $V_{\rm p}$. Considering that $M_{\rm Ni}$ can be robustly inferred from the radioactive tail based on the assumption of full $\gamma$-ray trapping at the nebular phase, our results provide an approach to quantify the effects of the $^{56}$Ni heating from observables.

Applying the above findings to a sample of SNe II, we find that the distribution of $M_{\rm Henv}$ estimated from the observed light curves is considerably broader than the ones predicted by single star models evolving with the standard stellar-wind prescription. This inconsistency suggests that a large fraction of SNe II experience substantial mass-loss before the onset of the core-collapse, pointing to missing ingredients that determine the mass-loss rate, either in the standard wind mass-loss or binary interaction, or both, to account for the diversity in $M_{\rm Henv}$, particularly at the low-mass end in the distribution of $M_{\rm Henv}$. 

However, it is important to address several limitations of this work. Firstly, we have assumed that the microphyiscs, such as convection and overshooting, are fixed throughout the study. Our approach is motivated by the range of $T_{\rm eff}$ observed for RSGs in the Galaxy. In practice, these factors may depend on $M_{\rm ZAMS}$, the evolution phases, or vary on a case-by-case basis. For example, by adjusting the mixing length in the hydrogen-rich envelope and the overshooting parameters, \citet{goldberg19} generated progenitors with large $M_{\rm Henv}$ but small $R$, which are missing in our model grid (see also \citealt{dessart13}). The $M_{\rm ZAMS}$-$R$ relation will also change if the progenitors suffer from radial pulsation prior to the explosion. Further, the mass-loss process is not self-consistently modeled. These factors potentially modify the structure of the hydrogen-rich envelope, and eventually affect the light curve at the plateau phase. Developing robust theory of convection would benefit from detailed 3D simulations of RSG (see \citealt{goldberg22a, goldberg22b} for recent progress). In the future, the advance in observational techniques will provide better constrains on the RSG mass-loss rates and the binary fractions of massive stars. A comprehensive analysis that includes all these factors will improve the accuracy of the results in this work.

Although the properties of the hydrogen-rich envelope are very sensitive to the mass-loss history, it is important to note that the nucleosynthesis products within the helium core are hardly affected by the stripping of the outer envelope, but primarily determined by the helium core mass (e.g., \citealt{takahashi23}). The nebular observation, during which the ejecta become transparent and the intermediate-mass elements are exposed, is an useful tool to constrain the properties of the material in the innermost region (\citealt{wheeler15, haynie23}). In particular, the strength of the oxygen emission [O {\sc I}] is considered as a reliable measurement of $M_{\rm ZAMS}$ of the progenitor from theoretical ground (\citealt{fransson89,maeda07,jerkstrand12,jerkstrand14,jerkstrand15,jerkstrand17,dessart20,dessart21a,dessart21b,dessart23}), which is further applied to samples of observational data (see, e.g., \citealt{hanin15,fang18,fang19,fang22,prentice19,terreran19,hiramatsu21}). By directly comparing the results obtained from nebular spectroscopy and light curve modeling, it becomes possible to establish a relationship between $M_{\rm ZAMS}$ and the properties of the envelope, thus linking the progenitors with the mass-loss histories they experienced prior to their explosion as SNe II (Fang et al., in preparation).

%\begin{acknowledgements}
%The authors thank Takashi Nagao for reading the manuscript and providing valuable comments. KM acknowledges support from the Japan Society for the Promotion of Science (JSPS) KAKENHI grant (JP20H00174). TJM is supported by the Grants-in-Aid for Scientific Research of the Japan Society for the Promotion of Science (JP20H00174, JP21K13966, JP21H04997). TM acknowledges support from the Hakubi project at Kyoto University.
%\end{acknowledgements}

\software{$\texttt{MESA}$ \citep{paxton11, paxton13, paxton15, paxton18, paxton19, mesa23}; $\texttt{SNEC}$ \citep{snec15}; SciPy \citep{scipy}; NumPy \citep{numpy}; Astropy \citep{astropy13,astropy18}; Matplotlib \citep{matplotlib}}

\appendix
\clearpage
\newpage

\setcounter{table}{0}
\renewcommand{\thetable}{A\arabic{table}}

\startlongtable
\centerwidetable
\begin{deluxetable*}{cccccccccc}
\label{tab:SNe sample}
\tablecaption{SNe II sample in this work.}
\tablehead{\colhead{Name}&$t_{\rm exp}$&$E\,{\rm (}B-V{\rm )}$&$\mu$&$M_{\rm Ni}$&$t_{\rm p}$&$t_{\rm p,0}$&$V_{\rm p}$&$M_{\rm Henv}$&Ref.\\
\colhead{}&(MJD)&(mag)&(mag)&($M_{\rm \odot}$)&(days)&(days)&(mag)&($M_{\rm \odot}$)&}
\startdata
1986L&46708&0.03&31.72\,(0.20)&0.058\,(0.046)&111.6\,(6.0)&85.40\,(15.05)&-17.18\,(0.20)&9.66\,(3.04)&a,b\\
1992af&48791&0.00&34.33\,(0.12)&0.052\,(0.040)&72.2\,(6.0)&45.44\,(15.31)&-17.08\,(0.12)&3.05\,(1.55)&a,b\\
1992ba&48888&0.02&31.07\,(0.30)&0.019\,(0.006)&125.05\,(8.0)&99.40\,(12.48)&-15.71\,(0.30)&7.49\,(2.21)&a,b,c\\
1995ad&49981&0.04&31.80\,(0.15)&0.029\,(0.014)&79.55\,(3.0)&57.62\,(10.49)&-16.42\,(0.15)&3.41\,(1.26)&d\\
1999em&51475&0.06&30.34\,(0.07)&0.054\,(0.011)&123.53\,(1.0)&91.45\,(7.69)&-16.62\,(0.07)&8.71\,(1.56)&a,b,c,e\\
1999gi&51518&0.19&30.34\,(0.14)&0.032\,(0.002)&121.98\,(3.1)&89.61\,(6.73)&-15.98\,(0.14)&6.60\,(1.24)&b,c,f\\
2001X&51963&0.07&31.59\,(0.11)&0.055\,(0.005)&114.51\,(5.0)&77.38\,(7.75)&-16.44\,(0.11)&5.94\,(1.35)&b,g,h\\
2002gw&52560&0.14&32.98\,(0.22)&0.024\,(0.006)&102.21\,(3.0)&77.29\,(8.80)&-15.98\,(0.22)&5.06\,(1.40)&a,b,i,j\\
2002hj&52563&0.10&34.91\,(0.15)&0.030\,(0.024)&101.55\,(7.0)&77.76\,(14.47)&-16.46\,(0.15)&6.17\,(2.14)&a,b,i\\
2002hx&52580&0.18&35.53\,(0.08)&0.066\,(0.010)&72.49\,(3.7)&42.19\,(6.73)&-16.96\,(0.08)&2.29\,(0.87)&a,b,i,j\\
2003B&52622&0.05&30.62\,(0.25)&0.006\,(0.02)&100.27\,(4.2)&83.61\,(7.36)&-14.78\,(0.25)&3.77\,(0.95)&a,b,i,j\\
2003T&52655&0.03&35.36\,(0.15)&0.046\,(0.011)&103.64\,(10.0)&65.93\,(14.32)&-16.21\,(0.15)&4.17\,(1.82)&b,k\\
2003Z&52665&0.03&31.70\,(0.60)&0.005\,(0.003)&120.40\,(4.5)&101.54\,(11.79)&-14.33\,(0.60)&4.81\,(1.76)&b,k\\
2003bn&52695&0.06&33.55\,(0.15)&0.026\,(0.020)&118.40\,(3.0)&94.30\,(14.05)&-16.26\,(0.15)&8.20\,(2.20)&a,b\\
2003bl&52700&0.02&34.07\,(0.30)&0.009\,(0.008)&104.74\,(3.0)&83.90\,(12.20)&-15.10\,(0.30)&4.32\,(1.27)&a,b,l\\
2003cx&52729&0.08&35.91\,(0.15)&0.032\,(0.025)&94.06\,(5.0)&68.62\,(14.06)&-16.52\,(0.15)&5.00\,(1.77)&a,b\\
2003fb&52779&0.37&34.05\,(0.13)&0.034\,(0.008)&95.70\,(4.0)&51.39\,(11.32)&-15.58\,(0.13)&2.05\,(1.02)&j,k\\
2003hd&52858&0.01&36.02\,(0.15)&0.036\,(0.004)&94.09\,(5.0)&72.86\,(7.00)&-16.72\,(0.15)&5.88\,(1.32)&a,b\\
2003hg&52866&0.06&33.65\,(0.16)&0.014\,(0.011)&123.85\,(5.0)&100.49\,(13.37)&-15.79\,(0.16)&7.71\,(1.94)&a,b,l\\
2003hk&52868&0.14&34.77\,(0.12)&0.028\,(0.007)&86.09\,(3.0)&74.10\,(5.42)&-17.22\,(0.12)&7.24\,(1.19)&j,k,m\\
2003hl&52869&0.06&32.16\,(0.10)&0.011\,(0.008)&135.60\,(5.0)&114.12\,(12.26)&-15.30\,(0.10)&8.20\,(1.68)&a,b,g\\
2003hn&52857&0.13&31.14\,(0.26)&0.032\,(0.005)&106.68\,(4.0)&85.90\,(7.43)&-16.58\,(0.26)&7.72\,(1.84)&a,b,c\\
2003iq&52920&0.06&32.16\,(0.10)&0.049\,(0.009)&95.46\,(2.0)&53.56\,(8.25)&-16.12\,(0.10)&2.65\,(0.96)&a,b,g\\
2004A&53012&0.18&30.87\,(0.26)&0.026\,(0.007)&118.19\,(2.0)&89.57\,(9.58)&-15.90\,(0.26)&6.52\,(1.74)&j,n,o\\
2004dj&53181&0.09&27.46\,(0.11)&0.013\,(0.004)&110.09\,(15.6)&93.47\,(16.53)&-15.80\,(0.11)&6.87\,(2.36)&j,p,q,r,s\\
2004ej&53232&0.14&33.1\,(0.21)&0.017\,(0.007)&105.53\,(4.2)&93.08\,(6.96)&-16.63\,(0.21)&9.09\,(1.67)&a,j\\
2004er&53272&0.13&33.83\,(0.15)&0.033\,(0.026)&150.59\,(2.0)&125.79\,(12.72)&-16.55\,(0.15)&15.64\,(3.04)&a,b\\
2004et&53270&0.41&28.36\,(0.09)&0.068\,(0.009)&123.50\,(4.0)&86.65\,(7.42)&-16.70\,(0.09)&8.10\,(1.51)&b,t,u\\
2004fx&53304&0.09&32.71\,(0.15)&0.017\,(0.007)&102.14\,(4.0)&74.34\,(11.67)&-15.43\,(0.15)&3.83\,(1.30)&a,b\\
2005J&53383&0.22&33.96\,(0.14)&0.059\,(0.045)&114.37\,(7.0)&87.18\,(16.19)&-17.21\,(0.14)&10.08\,(3.29)&a,b,j\\
2005ay&53450&0.04&30.68\,(0.21)&0.017\,(0.004)&114.35\,(1.8)&84.58\,(9.59)&-15.35\,(0.21)&4.75\,(1.27)&j,v\\
2005cs&53548&0.05&29.26\,(0.33)&0.002\,(0.001)&125.61\,(0.5)&116.31\,(4.38)&-14.65\,(0.33)&6.69\,(1.16)&a,b,j,v,w,x\\
2005dk&53600&0.00&34.01\,(0.14)&0.044\,(0.034)&99.76\,(6.0)&73.42\,(15.45)&-16.89\,(0.14)&6.52\,(2.38)&a,b\\
2005dx&53616&0.09&35.09\,(0.09)&0.010\,(0.004)&100.76\,(4.6)&85.32\,(8.34)&-15.54\,(0.09)&5.13\,(1.09)&a,b,j\\
2005dz&53620&0.07&34.44\,(0.15)&0.024\,(0.018)&112.51\,(4.0)&88.87\,(12.98)&-16.18\,(0.15)&7.10\,(1.91)&a,b\\
2006Y&53767&0.11&35.70\,(0.06)&0.044\,(0.032)&67.19\,(4.0)&41.36\,(14.27)&-16.88\,(0.06)&2.37\,(1.24)&a,b,y\\
2006ai&53782&0.11&34.01\,(0.14)&0.054\,(0.041)&71.32\,(5.0)&45.35\,(14.44)&-17.11\,(0.14)&3.00\,(1.46)&a,b,y\\
2006ee&53962&0.00&33.87\,(0.15)&0.020\,(0.015)&106.69\,(4.0)&83.70\,(12.12)&-15.99\,(0.15)&5.93\,(1.63)&a\\
2006ov&53974&0.02&30.50\,(0.95)&0.002\,(0.002)&120.84\,(6.0)&110.50\,(9.33)&-14.80\,(0.95)&6.86\,(3.11)&k\\
2007ab&54124&0.23&34.97\,(0.15)&0.048\,(0.038)&73.16\,(10.0)&47.32\,(17.07)&-16.96\,(0.15)&3.19\,(1.79)&a,b\\
2007it&54349&0.42&30.35\,(0.36)&0.108\,(0.033)&112.62\,(10.0)&76.24\,(17.99)&-17.03\,(0.36)&7.66\,(3.86)&a,b,j,z\\
2007od&54388&0.04&32.05\,(0.15)&0.020\,(0.010)&135.83\,(4.1)&125.92\,(6.23)&-17.28\,(0.15)&20.40\,(2.38)&aa\\
2007sq&54422&0.00&34.12\,(0.13)&0.004\,(0.003)&105.81\,(4.0)&88.29\,(10.78)&-14.41\,(0.13)&3.61\,(0.95)&a\\
2008K&54478&0.09&35.29\,(0.10)&0.027\,(0.012)&93.45\,(4.0)&74.88\,(9.05)&-16.55\,(0.10)&5.85\,(1.47)&a,b,j\\
2008M&54475&0.09&32.62\,(0.20)&0.027\,(0.011)&81.43\,(3.3)&61.33\,(9.11)&-16.43\,(0.20)&3.88\,(1.26)&a,b,j\\
2008W&54486&0.00&34.59\,(0.11)&0.019\,(0.015)&97.57\,(6.0)&74.37\,(13.60)&-15.99\,(0.11)&4.74\,(1.63)&a\\
2008aw&54518&0.14&33.21\,(0.15)&0.087\,(0.023)&77.93\,(4.0)&52.37\,(8.79)&-17.45\,(0.17)&4.16\,(1.49)&a,b,j\\
2008bk&54543&0.02&27.68\,(0.13)&0.007\,(0.001)&131.65\,(2.0)&112.97\,(5.57)&-14.73\,(0.13)&6.45\,(0.90)&ab,ac\\
2008bx&54576&0.02&32.99\,(0.80)&0.033\,(0.051)&87.88\,(4.0)&63.37\,(13.51)&-16.60\,(0.80)&4.58\,(2.25)&ad\\
2008ea&54646&0.12&33.77\,(0.16)&0.010\,(0.008)&93.94\,(8.0)&72.96\,(14.09)&-15.31\,(0.16)&3.58\,(1.32)&ad\\
2008ga&54712&0.00&33.99\,(0.14)&0.005\,(0.004)&87.60\,(4.0)&69.48\,(11.15)&-14.52\,(0.14)&2.41\,(0.80)&ad\\
2008if&54808&0.10&33.54\,(0.15)&0.057\,(0.045)&83.96\,(5.0)&58.12\,(13.73)&-17.20\,(0.15)&4.74\,(1.98)&a,b\\
2008in&54808&0.05&30.45\,(0.10)&0.004\,(0.001)&107.83\,(1.0)&97.34\,(4.45)&-15.07\,(0.10)&5.55\,(0.73)&a,b,ae,af\\
2009N&54848&0.05&31.67\,(0.11)&0.016\,(0.002)&108.67\,(1.2)&81.51\,(5.87)&-15.44\,(0.11)&4.57\,(0.87)&a,b,ag\\
2009at&54899&0.55&31.82\,(0.22)&0.018\,(0.005)&75.90\,(2.0)&62.34\,(5.70)&-16.59\,(0.22)&4.16\,(1.03)&j,ad\\
2009bw&54916&0.28&31.44\,(0.15)&0.023\,(0.002)&136.13\,(3.0)&119.91\,(5.77)&-16.55\,(0.15)&14.19\,(1.82)&a,b,ah\\
2009dd&54916&0.45&30.74\,(0.15)&0.060\,(0.014)&126.66\,(4.2)&94.94\,(12.04)&-16.03\,(0.15)&4.90\,(1.69)&d,j,ai\\
2009ib&55041&0.16&31.48\,(0.31)&0.045\,(0.008)&140.65\,(2.0)&91.20\,(11.58)&-15.75\,(0.31)&6.47\,(2.22)&j,aj\\
2009kr&55140&0.07&32.09\,(0.15)&0.008\,(0.001)&82.97\,(2.0)&70.98\,(4.85)&-15.84\,(0.15)&4.04\,(0.80)&b,ak\\
2009md&55162&0.12&31.66\,(0.15)&0.004\,(0.003)&117.86\,(8.0)&103.27\,(11.31)&-14.72\,(0.15)&5.52\,(1.30)&b,al\\
2011ef&55760&0.06&33.60\,(0.18)&0.050\,(0.032)&117.41\,(1.0)&91.94\,(13.58)&-16.80\,(0.18)&9.54\,(2.58)&ad\\
2012A&55932&0.04&29.96\,(0.15)&0.009\,(0.001)&106.94\,(2.0)&93.31\,(5.15)&-15.61\,(0.15)&6.19\,(0.96)&b,am\\
2012aw&56002&0.08&29.96\,(0.09)&0.050\,(0.006)&135.74\,(4.0)&105.28\,(7.22)&-16.60\,(0.09)&11.31\,(1.69)&b,k,af,an\\
2012ck&56064&0.08&36.30\,(0.05)&0.071\,(0.055)&76.28\,(2.0)&49.71\,(14.68)&-17.44\,(0.05)&4.02\,(1.73)&ad\\
2012ec&56143&0.14&31.32\,(0.15)&0.039\,(0.005)&106.94\,(5.0)&84.39\,(7.88)&-16.75\,(0.15)&7.87\,(1.68)&b,ao\\
2013K&56302&0.25&32.66\,(0.50)&0.012\,(0.010)&131.40\,(5.0)&111.40\,(14.35)&-15.69\,(0.50)&9.43\,(3.20)&ap\\
2013ab&56340&0.04&31.90\,(0.08)&0.064\,(0.003)&102.12\,(1.0)&66.34\,(5.08)&-16.68\,(0.08)&4.84\,(0.90)&aq\\
2013am&56372&0.65&30.54\,(0.40)&0.015\,(0.006)&108.92\,(2.0)&91.72\,(8.87)&-16.02\,(0.40)&7.19\,(2.02)&ap\\
2013bu&56400&0.08&30.79\,(0.08)&0.002\,(0.001)&102.47\,(4.5)&92.71\,(6.55)&-14.45\,(0.08)&4.01\,(0.74)&b,ar\\
2013by&56404&0.23&30.81\,(0.15)&0.032\,(0.004)&84.79\,(2.0)&72.98\,(4.87)&-17.39\,(0.15)&7.52\,(1.22)&b,as\\
2013ej&56496&0.06&29.79\,(0.20)&0.021\,(0.002)&100.82\,(1.0)&86.75\,(4.96)&-16.62\,(0.20)&7.84\,(1.30)&b,at,au\\
2013fs&56571&0.04&33.45\,(0.15)&0.054\,(0.001)&80.20\,(0.5)&53.84\,(5.26)&-16.91\,(0.15)&3.56\,(0.94)&b,av\\
2013hj&56637&0.10&32.25\,(0.15)&0.080\,(0.008)&102.53\,(1.5)&76.16\,(6.13)&-17.32\,(0.15)&8.02\,(1.58)&aw\\
LSQ13dpa&56643&0.04&35.08\,(0.15)&0.071\,(0.013)&129.02\,(2.0)&93.21\,(7.99)&-16.80\,(0.15)&9.68\,(1.97)&a,b\\
2014G&56668&0.21&31.90\,(0.15)&0.034\,(0.001)&87.12\,(1.0)&73.17\,(4.56)&-17.22\,(0.15)&7.06\,(1.16)&aw,ax\\
2014cx&56902&0.10&31.27\,(0.47)&0.056\,(0.008)&109.52\,(1.0)&71.49\,(12.94)&-16.47\,(0.47)&5.57\,(2.75)&ay\\
2014cy&56900&0.36&31.85\,(0.34)&0.027\,(0.006)&124.90\,(1.0)&104.23\,(7.91)&-16.43\,(0.34)&10.63\,(2.59)&az\\
2014dw&56958&0.22&32.46\,(0.15)&0.009\,(0.001)&91.90\,(10.0)&81.87\,(11.24)&-16.26\,(0.15)&6.21\,(1.74)&b\\
ASASSN-14dq&56841&0.07&33.26\,(0.15)&0.046\,(0.008)&100.99\,(5.5)&78.53\,(8.15)&-16.95\,(0.15)&7.37\,(1.71)&b\\
ASASSN-14gm&56901&0.10&31.74\,(0.15)&0.077\,(0.010)&110.57\,(1.5)&79.54\,(6.68)&-17.07\,(0.15)&7.91\,(1.57)&b\\
ASASSN-14ha&56910&0.01&29.53\,(0.50)&0.010\,(0.003)&136.50\,(1.5)&103.07\,(15.10)&-14.37\,(0.50)&5.06\,(2.08)&b\\
2015V&57112&0.03&31.63\,(0.22)&0.023\,(0.006)&116.31\,(4.3)&84.02\,(10.71)&-15.59\,(0.22)&5.17\,(1.48)&j,ad\\
2015an&57268&0.09&32.42\,(0.13)&0.021\,(0.010)&130.21\,(1.6)&114.06\,(7.12)&-16.48\,(0.13)&12.58\,(1.78)&ba\\
2015cz&57298&0.48&34.02\,(0.20)&0.070\,(0.010)&113.26\,(4.3)&90.35\,(7.62)&-17.38\,(0.20)&11.33\,(2.34)&az\\
2016B&57382&0.08&32.14\,(0.40)&0.082\,(0.019)&133.71\,(1.2)&93.61\,(13.65)&-16.82\,(0.40)&10.36\,(3.99)&bb\\
2016X&57406&0.04&30.91\,(0.43)&0.034\,(0.006)&94.97\,(0.6)&67.83\,(10.22)&-16.34\,(0.43)&4.74\,(1.92)&bc\\
2016gfy&57641&0.21&32.36\,(0.18)&0.033\,(0.003)&112.72\,(0.9)&83.54\,(6.15)&-17.17\,(0.18)&8.97\,(1.76)&bd\\
2017it&57747&0.03&36.48\,(0.12)&0.100\,(0.010)&109.20\,(1.0)&76.52\,(5.83)&-17.31\,(0.12)&8.02\,(1.45)&be\\
2017ahn&57792&0.26&32.59\,(0.43)&0.041\,(0.006)&56.17\,(0.5)&40.54\,(6.68)&-17.34\,(0.43)&2.55\,(1.15)&bf,bg\\
2017eaw&57886&0.41&29.18\,(0.20)&0.115\,(0.027)&117.08\,(1.0)&79.22\,(10.25)&-17.26\,(0.20)&8.53\,(2.43)&bh,bi,bj\\
2017gmr&57999&0.30&31.46\,(0.15)&0.142\,(0.031)&96.54\,(1.0)&69.31\,(7.65)&-17.91\,(0.15)&8.36\,(2.03)&bk,bl,bm\\
2018gj&58128&0.08&31.46\,(0.15)&0.026\,(0.007)&77.84\,(1.4)&59.15\,(6.39)&-16.47\,(0.16)&3.60\,(0.98)&bn\\
2018zd&58178&0.17&29.91\,(0.22)&0.009\,(0.001)&116.59\,(1.0)&104.78\,(4.86)&-15.92\,(0.22)&8.71\,(1.27)&bo,bp,bq,br\\
2018cuf&58292&0.14&33.10\,(0.30)&0.040\,(0.010)&111.24\,(1.0)&86.47\,(9.03)&-16.62\,(0.30)&8.06\,(2.25)&bs\\
2018hfm&58395&0.31&32.79\,(0.64)&0.015\,(0.005)&57.21\,(5.3)&48.03\,(7.23)&-17.40\,(0.64)&3.53\,(1.46)&bt\\
2018hwm&58425&0.02&33.58\,(0.19)&0.003\,(0.002)&144.41\,(1.0)&134.13\,(5.06)&-14.94\,(0.19)&9.67\,(1.13)&bu\\
2020jfo&58974&0.02&30.81\,(0.20)&0.018\,(0.007)&66.15\,(2.0)&49.21\,(7.41)&-16.18\,(0.20)&2.29\,(0.86)&bv,bw,bx,by\\
2021gmj&59293&0.06&31.42\,(0.20)&0.020\,(0.004)&106.24\,(1.0)&77.98\,(7.87)&-15.60\,(0.20)&4.48\,(1.16)&bz\\
2021yja&59465&0.10&31.85\,(0.45)&0.141\,(0.050)&124.26\,(1.5)&78.86\,(19.62)&-17.26\,(0.45)&9.28\,(5.06)&ca,cb,cc\\
\hline
\enddata
\tablecomments{The columns are (from left to right): SN name, date of explosion, extinction, distance module, nickle mass, plateau duration, plateau duration corrected for nickle heating, plateau magnitude, hydrogen-rich envelope mass, references. References: a:\citet{anderson14}; b:\citet{valenti16}; c:\citet{jones09}; d:\citet{inserra13}; e:\citet{leonard03}; f:\citet{leonard02}; g:\citet{poznanski09}; h:\citet{faran14}; i:\citet{galbany16}; j:\citet{iron_A}; k:\citet{spiro14}; l:\citet{olivares10};  m:\citet{faran14b}; n:\citet{guru08}; o:\citet{hendry06}; p:\citet{vinko06}; q:\citet{zhang06}; r:\citet{tsvetkov08}; s:\citet{vinko09}; t:\citet{utrobin09}; u:\citet{maguire10}; v:\citet{tsvetkov06}; w:\citet{takats06}; x:\citet{pastorello09}; y:\citet{hiramatsu21}; z:\citet{andrews11}; aa:\citet{inserra11}; ab:\citet{pignata13}; ac:\citet{vandyk12a}; ad:\citet{ucb19}; ae:\citet{roy11}; af:\citet{bose14}; ag:\citet{takats14}; ah:\citet{inserra12}; ai:\citet{hicken17}; aj:\citet{takats15}; ak:\citet{eliasrosa10}; al:\citet{fraser11}; am:\citet{tomasella13}; an:\citet{dallora14}; ao:\citet{barbarino15}; ap:\citet{tomasella18}; aq:\citet{bose15}; ar:\citet{kanbur03}; as:\citet{valenti15}; at:\citet{fraser14}; au:\citet{valenti14}; av:\citet{yaron17}; aw:\citet{bose16}; ax:\citet{terreran16}; ay:\citet{huang16}; az:\citet{dastidar21}; ba:\citet{dastidar19}; bb:\citet{dastidar19b}; bc:\citet{huang18}; bd:\citet{singh19};
be:\citet{afsariardchi19}; bf:\citet{tartaglia21}; bg:\citet{nagao21}; bh:\citet{tsvetkov18}; bi:\citet{buta19}; bj:\citet{szalai19}; bk:\citet{andrews19}; bl:\citet{nagao19}; bm:\citet{utrobin21}; bn:\citet{teja23}; bo:\citet{zhang20}; bp:\citet{hiramatsu21b}; bq:\citet{callis21}; br:\citet{tsvetkov22}; bs:\citet{dong21}; bt:\citet{zhang22}; bu:\citet{reguitti21}; bv:\citet{sollerman21}; bw:\citet{teja22}; bx:\citet{ailawadhi23}; by:\citet{kilpatrick23b}; bz:\citet{murai24}; ca:\citet{vasylyev22}; cb:\citet{kozyreva22}; cc:\citet{hosseinzadeh22}
}
\end{deluxetable*}

\clearpage
\newpage
%TBD: 2008gz(?), 2015ba?, 2015bs, 2018aoq, 2019va, 2019mhm? 2019edo, 2019nyk, 2023axu, 2023ixf 2022acko(?) 2023ufx(? Low metal)

{}
%\end{CJK*}
\end{document}